\documentclass[twocolumn,preprintnumbers,amsmath,amssymb,superscriptaddress,nofootinbib]{revtex4-1}

\usepackage{graphicx}
\usepackage{dcolumn}
\usepackage{bm}
\usepackage{tabularx}
\usepackage{makecell}
\usepackage{braket}
\usepackage{lipsum}
\usepackage{ulem}
\usepackage{amsmath,amssymb} 
\usepackage[export]{adjustbox}
\usepackage{float}
\usepackage{caption}
\usepackage{ragged2e}
\usepackage{subcaption}
\usepackage{enumitem}
\usepackage{xcolor}
\usepackage[colorlinks=true, linkcolor=blue, urlcolor=blue, citecolor=blue]{hyperref}

\makeatletter
\long\def\@makecaption#1#2{%
  \vskip\abovecaptionskip
  {\small
   \noindent
   \justifying
   \textbf{#1.} #2\par}
  \vskip\belowcaptionskip
}
\makeatother

\begin{document}
	\title{Importance and Science Outcomes from the first XSPECT/\textit{XPoSat} Workshop}
	
\author{Anuj Nandi}
\email{anuj@ursc.gov.in}
\affiliation{Space Astronomy Group, ISITE Campus, U. R. Rao Satellite Centre, ISRO, Bengaluru, 560037, India}

\author{Rwitika Chatterjee}
\email{rwitika@ursc.gov.in}
\affiliation{Space Astronomy Group, ISITE Campus, U. R. Rao Satellite Centre, ISRO, Bengaluru, 560037, India}

\author{V. P. Shyam Prakash}
\email{shyamvp151@gmail.com}
\affiliation{Space Astronomy Group, ISITE Campus, U. R. Rao Satellite Centre, ISRO, Bengaluru, 560037, India}
\affiliation{Department of Physics, University of Calicut, Kerala, 673635, India}

\author{Ankur Kushwaha}
\affiliation{Space Astronomy Group, ISITE Campus, U. R. Rao Satellite Centre, ISRO, Bengaluru, 560037, India}

\author{M. C. Ramadevi}
\affiliation{Science and Space Exploration Programme Office, U. R. Rao Satellite Centre, ISRO, Bengaluru, 560037, India}

\author{Kiran M. Jayasurya}
\affiliation{Space Astronomy Group, ISITE Campus, U. R. Rao Satellite Centre, ISRO, Bengaluru, 560037, India}

\author{Vivek K. Agrawal}
\affiliation{Space Astronomy Group, ISITE Campus, U. R. Rao Satellite Centre, ISRO, Bengaluru, 560037, India}

\author{M. Varun}
\affiliation{Department of Physics and Electronics, CHRIST (Deemed to be University), Bangalore, 560027.}

\author{Karan Akbari}
\affiliation{St. Xavier's College (Empowered Autonomous), Mumbai, 400001}

\author{Arya Sudhakaran}
\affiliation{Manipal Centre for Natural Sciences, Manipal Academy of Higher Education, Manipal, Karnataka, India}

\author{Daneshwar Bhandari}
\affiliation{Pt. Ravishankar Shukla University Raipur-492010, Chhattisgarh, India}

\author{Vishal Kale}
\affiliation{School of Physical Sciences, S. R. T. M. University, Nanded -431606, Maharashtra, India}

\author{Vishal Jadoliya}
\affiliation{Department of Physics, Indian Institute of Technology Hyderabad, India}

\author{Suchismito Chattopadhyay}
\affiliation{Inter-University Centre for Astronomy and Astrophysics (IUCAA), Pune University Campus, Ganeshkhind, Pune, 411007, India}

\author{Sakshi Maurya}
\affiliation{Department of Physics, Institute of Science, Banaras Hindu University, Uttar Pradesh, India}

\author{Swasthik Visakh S}
\affiliation{Indian Institute of Space Science and Technology, Kerala, India}

\author{Debasish Krishnatreya}
\affiliation{Department of Physics, Tezpur University, Assam, India}

\author{M Dhamodhar Reddy}
\affiliation{Department of Astronomy, University college of science, Osmania University, Telangana, India}

\author{Akash Agarwal}
\affiliation{Astronomy and Astrophysics Group, Raman Research Institute, Bangalore, 560080, India}

\author{Giridharan L.}
\affiliation{Department of Physics and Electronics, CHRIST (Deemed to be University), Bangalore, 560027.}

\author{Meghamani Halder}
\affiliation{Department of physics, University of North Bengal, Darjeeling-734013, West Bengal, India}

\author{Juris N. J.}
\affiliation{Department of physics, St. Thomas College, Ranni 689673, India}

\author{Athira Mohanan}
\affiliation{Department of Physics, Central University of Karnataka, Kadaganchi 585367, India}

\author{P. Majumder}
\affiliation{Department of Physics, Rishi Bankim Chandra College, Naihati, West Bengal, 743165, India.}

\author{Arbind Pradhan}
\affiliation{Department of Applied Sciences, Tezpur University, Assam, India}

\author{Deblina Lahiri}
\affiliation{Department of Astronomy, University college of science, Osmania University, Telangana, India}

\author{Abhishek Jhala}
\affiliation{Manipal Centre for Natural Sciences, Manipal Academy of Higher Education, Manipal, Karnataka, India}

\author{Garima Tyagi}
\affiliation{Department of Physics, Jamia Millia Islamia, Jamia Nagar, New Delhi, 110025, India}

\author{Abhisek Tamang}
\affiliation{Astronomy and Astrophysics Group, Raman Research Institute, Bangalore, 560080, India}
\affiliation{Indian Institute of Science, C V Raman Avenue, Bangalore, Karnataka, 560012, India}

\author{Akshara S B}
\affiliation{School of Advanced Sciences, Vellore Institute of Technology, Vellore, Tamil Nadu, 632014, India}

\author{P Aromal}
\affiliation{Department of Astronomy, Astrophysics, and Space Engineering, Indian Institute of Technology Indore, India}

\author{Sreetama Das Choudhury}
\affiliation{Indian Institute of Technology Guwahati, Guwahati 781039, India}

\author{Sunirmal Rana}
\affiliation{Midnapore City College, Kuturiya, Bhadutala, Midnapore, West Bengal, 721129, India}

\author{Ajith Balu}
\affiliation{Astronomy and Astrophysics Group, Raman Research Institute, Bangalore, 560080, India}

\author{Mary Bosco}
\affiliation{Department of Physics and Electronics, CHRIST (Deemed to be University), Bangalore, 560027.}

\author{Jeyaraman Ilangovan}
\affiliation{Department of Earth and Space Sciences, Indian Institute of Space Science and Technology, Thiruvananthapuram 695547, Kerala, India}

\author{Shivani Chaudhary}
\affiliation{Department of Physics, Banasthali Vidyapith, Rajasthan, 304022, India}

\author{Bhranti Rao}
\affiliation{Department of Physics, University of Mumbai, Mumbai 400068, Maharashtra, India}

\author{Biyas Chowdhury}
\affiliation{Department of Physical Sciences, Indian Institute of Science Education and Research, Kolkata, Mohanpur 741252, West Bengal, India}

\author{Joysankar Majumdar}
\affiliation{Department of Physics, Institute of Science, Banaras Hindu University, Varanasi, 221005, India}

\author{Sanjeeva Rao Prattipati}
\affiliation{Department of Physics, Indian Institute of Technology Jodhpur, Karwar, Jodhpur, 342037, Rajasthan, India}

\author{Subhasish Das}
\affiliation{Department of Pure and Applied Physics, Guru Ghasidas Vishwavidyalaya, Bilaspur, India, 495009}

\author{Sandip Naskar}
\affiliation{Indian Institute of Technology Guwahati, Guwahati 781039, India}

\author{Swapnil Singh}
\affiliation{Space Astronomy Group, ISITE Campus, U. R. Rao Satellite Centre, ISRO, Bengaluru, 560037, India}

\author{Vaishali S}
\affiliation{Space Astronomy Group, ISITE Campus, U. R. Rao Satellite Centre, ISRO, Bengaluru, 560037, India}

\author{Koushal Vadodariya}
\affiliation{Space Astronomy Group, ISITE Campus, U. R. Rao Satellite Centre, ISRO, Bengaluru, 560037, India}

\author{Radhakrishna Vatedka}
\affiliation{Space Astronomy Group, ISITE Campus, U. R. Rao Satellite Centre, ISRO, Bengaluru, 560037, India}

\author{Anurag Tyagi}
\affiliation{Space Astronomy Group, ISITE Campus, U. R. Rao Satellite Centre, ISRO, Bengaluru, 560037, India}

\author{Sachin Narang}
\affiliation{ISRO Telemetry Tracking and Command Network, ISRO, Bengaluru, 560058, India}

\author{Prapti Mittal}
\affiliation{ISRO Telemetry Tracking and Command Network, ISRO, Bengaluru, 560058, India}

\author{Madhu K V}
\affiliation{Missions planning and operations group, U. R. Rao Satellite Centre, ISRO, Bengaluru, 560037, India}

\author{Brindavan Mahto}
\affiliation{Project program office, U. R. Rao Satellite Centre, ISRO, Bengaluru, 560037, India}

\date{\hfill \today}
	
%%%%%%%%%%%%%%%%%%%%%%%%%%%%%%%%%%%%%%%%%
	
\begin{abstract}

\noindent
\textbf{Abstract:} This paper summarizes the science outcomes of the first Workshop on data analysis \& modeling using observations from the XSPECT payload onboard \textit{XPoSat} mission, which brought together early career researchers and domain experts to explore the scientific capabilities of the instrument through scientific lectures and hands on analyses. 
Participants engaged in end-to-end XSPECT data handling, including calibration, spectral modeling, and timing analysis, performed on seven diverse astrophysical sources encompassing Neutron star Low Mass X-ray Binaries (NS-LMXBs), Pulsars, and Black hole X-ray Binaries (BH-XRBs), highlighting the instrument’s diagnostic potential and capabilities. These sources were observed during the first year of XSPECT observations, and the data along with the necessary software developed in-house, was provided to the participants,  making them the first to explore XSPECT data beyond the instrument development team.
For NS-LMXBs, Aql X-1 exhibited a classical Type-I X-ray burst, allowing constraints on stellar radius through spectral fitting. Sco X-1, observed across its complete Z-track, revealed systematic spectral evolution driven by accretion rate fluctuations and disk-corona coupling, while Cir X-1 displayed orbital phase dependent transitions between hard and soft states, reflecting change in accretion geometry. In accretion powered pulsars, GX 301–2 showed a double peaked, energy dependent pulse profile and strong iron fluorescence lines due to stellar wind reprocessing, whereas Vela X-1 exhibited orbital phase dependent absorption and steady coronal temperatures consistent with traversal through a critical accretion process. Among BH-XRBs, Cyg X–1 transitioned from a hard to soft-intermediate state with increasing disk contribution and spectral softening, while Cyg X-3 remained in the intermediate state with multiple emission lines originating from a clumpy stellar wind.
The workshop outcomes highlight the scientific promise of XSPECT data and emphasize the importance of collaborative training in maximizing the science return from XSPECT and future Indian space astronomy missions.
\end{abstract}

\maketitle

\section{Introduction}
X-ray astronomy (\cite{1962PhRvL...9..439G}) plays a crucial role in advancing our understanding of compact objects such as white dwarfs, neutron stars, and black holes (\cite{1983bhwd.book.....S},\cite{1973blho.conf..343N}), which are typically invisible in optical wavebands. A key class of systems that serve as natural laboratories for studying these objects are X-ray binaries (\cite{1973A&A....24..337S}, \cite{2006csxs.book.....L}), consisting of an accreting compact object in a gravitationally bound system with a companion `normal' star. Observations of X-ray binaries in the low energy X-ray band ($\sim 0.1-10$~keV) are particularly important, as this range captures emissions originating from the accretion disks surrounding these dense cosmic objects. Timing analysis techniques are used to trace rapid variability,
Quasi-periodic Oscillations (QPOs), and pulsations that reveal the object’s spin, magnetic field, and accretion dynamics (\cite{1989ARA&A..27..517V}, \cite{2006ARA&A..44...49R}, \cite{2012A&A...542A..56N}).
Complementarily, spectral studies in this energy range provide insights into the temperature, structure, composition, and
relativistic effects shaping the emitted radiation. Thus, by combining timing and spectral studies of X-ray binaries, especially at low energies, astronomers can uncover crucial insights into the nature, structure, and fundamental characteristics of compact objects, paving the way for a deeper understanding of extreme astrophysical phenomena.

X-ray binaries (XRBs) are broadly categorized into Low-Mass X-ray Binaries (LMXBs) and High-Mass X-ray Binaries (HMXBs), based on the mass and evolutionary stage of the companion star (\cite{2002apa..book.....F}). In LMXBs, mass transfer from a low mass companion star to the compact object occurs via Roche lobe overflow through the inner Lagrange point, leading to the formation of an accretion disk that radiates strongly in X-rays (\cite{2006csxs.book..623T}). In contrast, HMXBs primarily accrete matter from the strong stellar winds of their massive companion stars, producing a smaller (or sometimes, transient) accretion disk
that also emits in X-rays.

LMXBs with neutron star as the accretor (NS-LMXBs) are characterized by a low spin period and a relatively lower surface magnetic field compared to other binary systems. 
Further, these sources are classified into `Z' and `Atoll' sources, on the basis of the patterns they trace in their Hardness-Intensity Diagram (HID) and Color-Color Diagram (CCD) (\cite{1989A&A...225...79H}). An `Atoll' source traces a `banana' and `island' state in its HID, whereas `Z' sources trace a Z-like pattern. These patterns are believed to be caused primarily due to changes in the mass accretion rate (\cite{1989A&A...225...79H}, \cite{Homan_2007}).
These sources trace a wide range of luminosities in different spectral states. The Eddington luminosity is the maximum luminosity at which radiation pressure balances gravitational attraction, setting an upper limit on stable accretion onto a compact object.
For example, for a $\sim$ 1.7 M$_{\odot}$ neutron star, the Eddington luminosity L$_{\mathrm{Edd}}$ is $\sim 10^{38}\mathrm{erg\ s^{-1}}$ \cite{1989A&A...225...79H}. 
Z-sources are highly luminous sources with L$\sim 0.1 - 1$ L$_{\mathrm{Edd}}$, whereas atoll sources have luminosities in the range of $0.01 - 0.1$ L$_{\mathrm{Edd}}$.

HMXBs are binary systems comprising of a compact object accreting material from a massive early-type (O or B spectral class) companion star (\cite{lewin1995book}). These systems are characterized by their strong and often variable X-ray emission, which arises from the conversion of gravitational potential energy into radiation as stellar wind or disk material is accreted onto the compact object (\cite{frank2002accretion}). In many HMXBs, particularly those hosting a neutron star, the X-ray emission is modulated by the neutron star spin, producing coherent pulsations (\cite{1971ApJ...167L..67G},\cite{nagase1989accretion}), and hence such systems are also called `pulsars'. The study of HMXBs provides key insights into the physics of accretion under extreme magnetic fields, stellar wind interactions, and binary evolution, and contributes to our broader understanding of compact object populations in the Galaxy (\cite{tauris2006formation}).

XRBs that harbor a black-hole as the accretor are termed as Black hole X-ray binary systems (BH-XRBs). These systems are transient in nature, and often show outburst signatures (\cite{2018JApA...39....5S}).
During an outburst, the X-ray flux can reach several orders of magnitude that of the quiescent state \cite{Remillard_2006}.
Sources remain in the quiescent / low flux state for an extended period, and occasionally go into outbursts that can last for several months.
The outbursts observed in BH-XRB systems are often associated with spectral state transitions (\cite{10.1143/PTPS.155.99}, \cite{Done_2007}, \cite{1995ApJ...455..623C}, \cite{2015ApJ...807..108I}).
The source remains in the Low/Hard State (LHS) before the onset of the burst and transits to Intermediate state (IMS) during the rise of the outburst.
It is in the High/Soft state (HSS) during the peak of the outburst.
This state is dominated by thermal emission processes following which the source enters the Very High State (VHS) also called the Steep Power Law (SPL) state (though not frequent) and eventually traces back to the LHS towards the end of the outburst.
The source typically traces a `Q' pattern in the Hardness Intensity Diagram (HID) (\cite{Homan2001}, \cite{homan_belloni}, \cite{2004MNRAS.355.1105F}, \cite{2012A&A...542A..56N}, \cite{2018JApA...39....5S}, \cite{2021MNRAS.508.2447B}) as the source goes through different spectral states.
The relative contribution of the flux from thermal and non-thermal processes vary during the spectral evolution of the source during the outburst.
(\cite{2006ARA&A..44...49R}).

XRBs have been studied by various X-ray telescopes since many decades, mainly using imaging, spectroscopic and timing techniques. In recent years, polarimetry has also emerged as a useful tool to probe the physics of these systems. Long-term monitoring of these sources with good spectral and temporal resolution, and high count rate handling capacity, can help us understand these sources more deeply. For instance, continuous observations can detect state transitions in binaries and the track the changes in accretion torque in pulsars.

 X-ray SPECtroscopy and Timing (XSPECT) instrument (\cite{2013ASInC...9...96S},\cite{rkrish2025},\cite{2025arXiv250609918C}) was developed with these objectives in mind, to attempt to answer some key science questions and address the gaps in understanding. XSPECT is a soft X-ray spectrometer with X-ray binaries as its prime observational target. The main scientific objectives of XSPECT includes study of soft excess in X-ray pulsars and understanding its nature and origins, long-term spin period and pulse profile studies, studying state changes in NS-LMXBs and the nature of soft thermal component, search for low frequency QPOs (LFQPOs) in NS and BH-XRBs in the soft X-ray band, measuring the spin and mass of the black holes through modeling of continuum and iron line profiles.

In addition to spectro-temporal studies, another instrument POLIX (POLarimeter Instrument in X-rays) (\cite{2022cosp...44.1853P}), the first medium energy ($8-30$ keV) X-ray polarimeter, is flown on-board \textit{XPoSat}\footnote{\href{https://www.isro.gov.in/XPoSat.html}{https://www.isro.gov.in/XPoSat.html}}. Combining polarimetry with spectral and temporal information can provide a more holistic view of such systems. The \textit{XPoSat} mission was envisaged to carry these two instruments to a low earth orbit to study bright celestial X-ray sources. 
\textit{XPoSat} (\cite{2025ExA....59...17S}) was launched on 2024~January~1 from Sriharikota, India, and placed into orbit at an altitude of 650~km and inclination of 6$^{\circ}$, similar to the orbit of \textit{AstroSat} (\cite{2006AdSpR..38.2989A}).

After completing the performance verification (PV) phase, \textit{XPoSat} commenced science operations on 2024~March~9. As of this writing (June 2026),
XSPECT has observed 41 sources across various source
categories (17 NS-LMXB, 10 pulsars, 4 BH-XRBs, 1 mag-
netar, and 9 AGNs) over the last 30 months. To have a fruitful usage of this significant amount of data, unlock the full scientific capabilities of XSPECT, and to broaden the user base, it was envisaged to provide hands-on training to a group of students on the utilization of XSPECT data. 

\begin{figure*}
    \centering
    \includegraphics[width=0.49\linewidth]{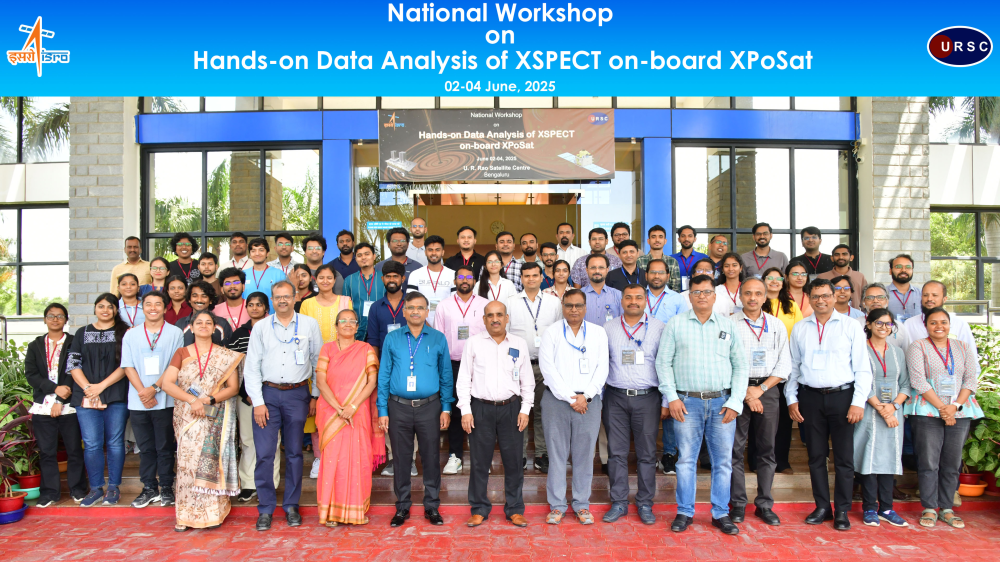}
    \includegraphics[width=0.49\linewidth]{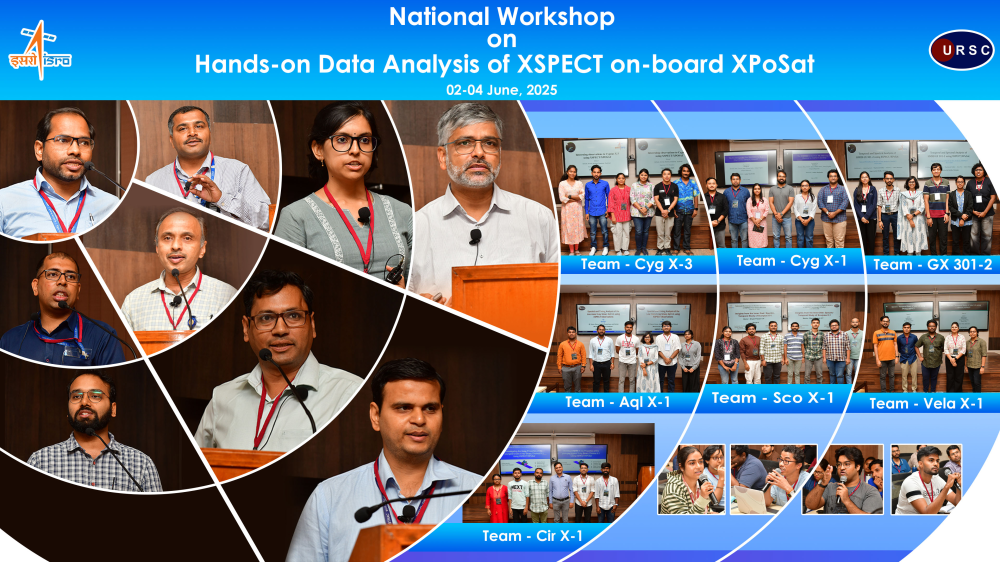}
    \caption{(Left) Participants and dignitaries of the workshop captured in a group photograph. (Right) Collage highlighting the scientific sessions, presentations, and collaborative project teams formed during the event.}
    \label{Group_hoto}
\end{figure*}

In this spirit, a Workshop was conducted during the period 2025~June~4~to~6. The primary objective of this workshop was to bring together young researchers, students, and experts from a diverse range of institutes to engage in discussions on the scientific capabilities of XSPECT. It provided an invaluable opportunity for participants to gain hands-on experience with the data and software associated with the payload, allowing them to explore its full potential in real-world applications. In Figure~\ref{Group_hoto}, a group photo captured during the workshop is displayed.

The remaining paper is organized as follows: in Section 2, we describe the XSPECT instrument in brief, Section 3 brings out the motivation behind organizing this workshop, Section 4 describes the sources explored by the workshop participants, Section 5 discusses the data reduction and processing aspects, Section 6 and 7 describe the data analysis as well as results with discussion, and finally, we conclude in Section 8.

\section{About XSPECT}
\begin{figure*}[ht]
    \centering
    \includegraphics[width=0.95\linewidth]{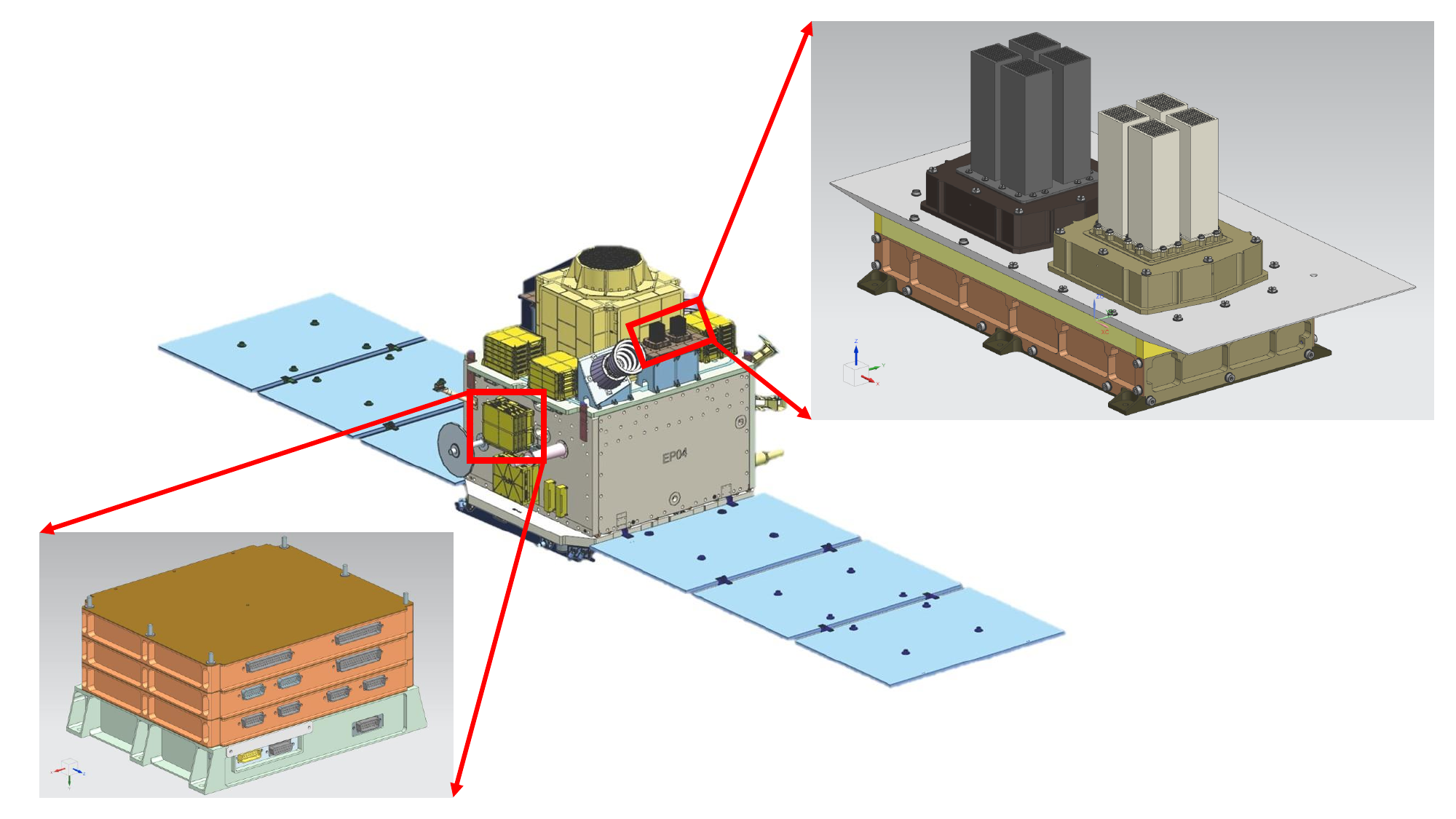}
    \caption{\textit{XPoSat} spacecraft in deployed configuration (centre) with zoomed in view of Electronics (left-bottom) and one detector package (right-top) of XSPECT payload. Another detector package with similar configuration is placed on the other side of POLIX.}
    \label{XSPECT_image}
\end{figure*}

XSPECT is a collimated X-ray telescope, using Swept Charge Devices (SCDs; \cite{2001NIMPA.458..568L}, \cite{2008SPIE.7021E..17H}) as its detectors. SCDs have previously been flown on Chandrayaan-1 (C1XS; \cite{2009P&SS...57..717G}), Chandrayaan-2 (CLASS; \cite{2011LPI....42.1708R}), and Insight-HXMT (LE Telescope; \cite{2020SCPMA..6349505C}). The instrument is sensitive in the soft X-ray band, and provides moderate energy and timing resolution. XSPECT (see Figure \ref{XSPECT_image}) consists of sixteen SCDs, having collimators with two kinds of fields of view (FOV), and also includes a blocked device (covered with a 500~$\mu$m sheet of Tantalum) to estimate the local particle background (\cite{rkrish2025},\cite{2025arXiv250609918C}). The instrument specifications are summarized in Table \ref{XSPECT_specifications}. 

\begin{table}[ht]
\caption{Specifications of XSPECT instrument}
    \centering
    \begin{tabular}{ll}
    \hline
    Parameter & Specification \\
    \hline
    Detectors   &  16 SCDs (1 blocked) \\
    Energy range   & 0.8 - 15.0 keV \\
    Energy resolution  & $\sim180$ eV at 6 keV \\
    Timing resolution   & 1 ms \\
    Effective area   & $>$ 30 cm$^2$ at 6 keV (all events) \\
    FOV   & $\sim 3^\circ \times 3^\circ$ and $\sim 2^\circ \times 2^\circ$\\
    \hline
    \end{tabular}
    \label{XSPECT_specifications}
\end{table}

XSPECT instrument was calibrated both on ground as well as on-board, to have a good understanding of the instrument performance as well as to validate its capabilities to carry out scientific studies. As part of ground tests, the instrument underwent thermovacuum cycles, as well as characterization of its gain and noise performance as a function of temperature (see \cite{2025arXiv250609918C} for details). 
%These two parameters are crucial for ... (\textcolor{red}{rephrase}). 
In addition, we also characterized its spectral response by carrying out monochromatic experiments at the Indus beamlines, \href{https://www.rrcat.gov.in/index_eng.html}{RRCAT, Indore}\footnote{\href{https://www.rrcat.gov.in/index_eng.html}{https://www.rrcat.gov.in/index\_eng.html}}. 

\begin{figure}[ht]
    \centering
    \includegraphics[scale=0.3, trim=6cm 1.8cm 10cm 4cm, clip=true]{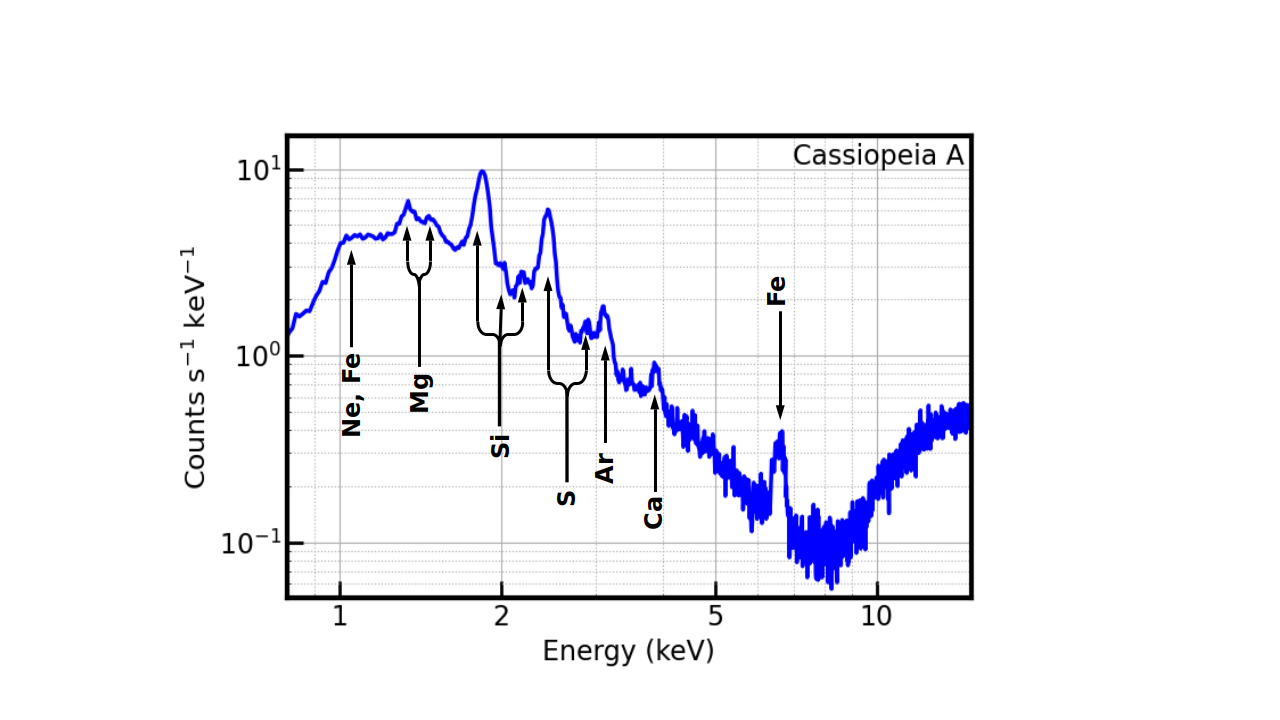}
    \caption{XSPECT first light: the spectrum of supernova remnant Cassiopeia~A observed by XSPECT, showing emission lines of several elements such as Mg, Si, S, Ar, Ca and Fe. Spectral distribution is without background subtraction.}
    \label{CasA_spectrum}
\end{figure}

After launch, XSPECT saw first light from supernova remnant (SNR) Cas~A (see Figure~\ref{CasA_spectrum}), which is a well-known calibration source in the soft X-ray band due to the presence of several emission lines in the energy band of interest. The observations of this source, as well as those of SNR Tycho, were used to validate the gain and spectral performance of each detector determined from ground experiments. Emission lines of several elements such as Mg (1.36 keV and 1.51 keV), Si (1.85 keV, 1.99 keV and 2.19 keV), S (2.43 keV and 2.60 keV), Ar (3.10 keV), Ca (3.87 keV), and Fe (6.60 keV) (\cite{1994PASJ...46L.151H}) were identified in the acquired spectra of these two SNRs.
For information regarding the design, calibration and performance, readers are referred to \cite{2025arXiv250609918C} and \cite{rkrish2025}

Following this, Crab observations were carried out to determine the alignment of each individual collimator with the spacecraft pointing axis. Although this was measured on ground, launch stresses and vibrations may have changed these values slightly. These observations are also used to fine-tune the detector effective area, validate the response, as well as verify the timing capabilities of the instrument. Additionally, blank sky observations were carried out to estimate the (instrument, as well as X-ray) background, which is crucial for utilizing data from non-imaging instruments with relatively large FOVs. 

%A dedicated Python-based software package was developed to process and analyze XSPECT data, which includes data reduction and screening, light curve and spectra generation, background generation, and other useful utilities. The response and background files were also generated to enable spectral analysis (rephrase). The data, along with the software and other files will be hosted at ISSDC (PRADAN link) for utilization by the community. In addition, being a proposal-driven mission, \textit{XPoSat} will accept/encourages proposals from the science community.

%\begin{itemize}
%    \item contemporary instruments and uniqueness of xspect
%\end{itemize}

\section{Motivation of the Workshop}

\begin{table*}
\renewcommand{\arraystretch}{1.4}
\centering
    \caption{Log of XSPECT observations of the sources provided during the workshop. The proposal IDs, dates, days of observations and the effective exposure in ks for each day are given in column 3, 4, 5 and 6 respectively.}
    \label{Source_observation_log}
\begin{tabular}{cccccc}
\hline
\hline
Sl. No. & Source Name & Proposal ID & Date (UTC) & Days of Observation & Effective Exposure (ks) \\
\hline
\multicolumn{6}{c}{NS-LMXBs} \\
\hline
1 & Aql X-1  & T24\_0009 & 2024-10-06  &   &  16.5  \\
&   & T24\_0009 & 2024-10-07  & 3  &  16.7  \\
&   & T24\_0009 & 2024-10-08  &   &  15.9  \\
\hline
2 & Sco X-1  &  G01\_0006  & 2024-08-06 &    & 12.4  \\
&   &  G01\_0006 & 2024-08-07 &  4  & 12.6  \\
&   & G01\_0006 & 2024-08-16 &    & 13.6  \\
&   &  G01\_0006 & 2024-08-17 &    & 13.2  \\
\hline
3 & Cir X-1      &  T24\_0008 & 2024-07-28 &  & 15.1\\  
&  & T24\_0008 & 2024-07-29 & 4 & 15.3 \\
&  & T24\_0008 & 2024-07-30 &  & 15.3 \\
&  & T24\_0008 & 2024-07-31 &  & 15.1  \\
\hline
\multicolumn{6}{c}{Pulsars} \\
\hline
4 &    GX 301-2  & G01\_0004 & 2024-04-12   &    &  21.8 \\
&  & G01\_0004 & 2024-04-13   &  3  &  21.8 \\
&  &  G01\_0004 & 2024-04-14   &    &  22.2 \\
\hline
5  & Vela X-1 &  G01\_0001 & 2024-04-07 &   & 18.4  \\
   &  &  G01\_0001 &  2024-04-08 & 3  &  18.0  \\
   &  &  G01\_0001 & 2024-04-09 &   &  18.6  \\
\hline
\multicolumn{6}{c}{BH-XRBs} \\
\hline
6  & Cyg X-1  & G02\_0004 & 2024-09-09  &  &  16.8  \\
  &   & G02\_0004 & 2024-09-10   &  4  &  17.0     \\
  &   & G02\_0004 & 2024-09-12   &    &   17.2    \\
  &   & G02\_0004 & 2024-09-13   &    &   17.0    \\
\hline
7  & Cyg X-3 & T24\_0011 & 2024-11-05 &  & 12.0 \\
   &         & T24\_0011 & 2024-11-06 & 3  & 11.5 \\
   &         & T24\_0011 & 2024-11-07 &   & 10.0 \\
\hline
\hline
\end{tabular}
\end{table*}

This workshop brought together young researchers and experts in the field of Astronomy to discuss the science capabilities of the XSPECT payload. The participants were grouped into seven teams, and each team was guided by the instrument team members through practical steps in analyzing XSPECT data, emphasizing both spectral and timing analysis techniques. 
In this way, this workshop was unique as it was a perfect blend of theoretical and practical experience of the data analysis of XSPECT. 
Lectures centered around the instrument’s perspective of the payloads provided valuable insights into both scientific and functional aspects of the instruments.
During the practical session, participants received training on the extraction and reduction of XSPECT data utilizing standard pipelines. 
Additionally, they acquired skills in extracting Level-2 data products, including light curves and spectra.
The most important part of the workshop was the group project, in which students and researchers from different parts of the country collaborated utilizing the data from
XSPECT to understand and explore the interesting science of compact objects.
The snippet of presentations by each group and the lecture session by instrument team is shown in the right panel of Figure \ref{Group_hoto}.

What made this particular workshop unique was the opportunity it offered to participants to work directly with XSPECT data from the \textit{XPoSat} mission, even before its public release. 
This was the first instance where actual observational data were made available to students and early-career researchers, allowing them to explore, analyze, and interpret it using the XSPECT data analysis tools. 
By facilitating direct interactions with the software tools, the workshop not only promoted a deeper understanding of XSPECT’s technical functions but also encouraged participants to provide critical feedback.
This feedback was valuable for identifying areas for improvement, allowing for iterative updates and refinements to the software, ultimately making it more user-friendly and accessible for researchers.

The sources identified for the guided projects in the workshop include various types of X-ray sources, such as NS-LMXBs, BH-XRBs, and Pulsars.
The details of the XSPECT observations along with brief characteristics of the sources are given in Sections \ref{sources} and \ref{obs_data_red}

\section{Sources: NS-LMXBs, Pulsars and BH-XRBs}\label{sources}

Seven X-ray sources from three categories (NS-LMXBs: Aql X-1, Sco X-1, Cir X-1; X-ray Pulsars: Vela X-1 and GX 301-2; BH-XRBs: Cyg X-1 and Cyg X-3) were considered, for seven projects identified for this workshop. The details of the sources and observation are provided in the following subsections. 

\subsection{NS-LMXBs}
\begin{itemize}
    \item Aql X-1 is classified as an `Atoll' source and consists of a neutron star accreting matter from a K-type low-mass companion star via Roche-lobe overflow (\cite{1991A&A...251L..11C}). The system is situated at an estimated distance of $\sim 6$~kpc. The system is known to show nearly annual outbursts (\cite{2013MNRAS.432.1695C}).
    Aql X-1 gained particular significance with the detection of a coherent pulsation at 550.27\,Hz during a short interval in the 1998 outburst by RXTE (\cite{2008ApJ...674L..41C}, \cite{Kocab_y_k_2025}). Detailed spectral analysis of this event revealed a soft X-ray excess during the pulse-on phase, attributed to a localized hotspot on the neutron star surface with a temperature of $\sim$1.65\,keV and a radius of $\sim$1.65\,km (\cite{Kocab_y_k_2025}). 
    
    \item Scorpius X-1 (hereafter, Sco X-1) was first discovered in 1962 as the brightest and persistent X-ray source in the galactic plane (\cite{1962PhRvL...9..439G}), located at a distance $\sim$ 2.8 kpc (\cite{Bradshaw_1999}). It emits near the Eddington limit ($\mathrm{L}_{\mathrm{Edd}}$) (\cite{1989A&A...225...79H}).  It is a prototype NS-LMXB system with a K-type companion star and an orbital period of 18.9 h (\cite{Gottlieb1975}). The mass of the companion star has been calculated to be lying in between $\rm \, 0.28 \, \mathrm{M}_{\odot} < \mathrm{M} < 0.70 \, \mathrm{M}_{\odot}$ (\cite{MataSanchez2015}). The centroid frequency of the QPO varies from 4.5-7.0 Hz in NB, while in FB the frequency varies from $6-25$ Hz (\cite{Casella2006}). Moreover, kilo-hertz (kHz) QPOs with lower kHz QPOs around 600 Hz and upper kHz QPOs around $875-1000$ Hz are also observed in this source (\cite{1997ApJ...481L..97V}).

    \item Cir X-1 is located in the constellation Circinus, in the southern sky with RA=230.170034 and DEC=-57.166724 \cite{2013wise.rept....1C}. Cir X-1 is a weakly magnetized Neutron Star X-ray Binary characterized by an eccentric ($e \sim 0.45$) $\sim$16.5 day orbit (\cite{1976ApJ...208L..71K},\cite{2020ApJ...891..150S}). The discovery of the natal supernova remnant of Cir X-1 makes it the youngest known XRB. \cite{2013ApJ...779..171H} and \cite{2006ApJ...653.1435B} observed twin-peak QPO in Cir X-1 and reported the inferred mass of $2.2 \pm 0.2 \mathrm{M}_\odot$ neglecting the rotation of neutron star.
\end{itemize}

\subsection{NS-HMXBs - Pulsars}
\begin{itemize}
    \item Vela X-1 is an HMXB system with a B0.5Ib optical companion, having a mass of $\sim23\,\mathrm{M}_\odot$ and a radius of $\sim30\,\mathrm{R}_\odot$ (\cite{vanKerkwijk1995b}). Accretor is a neutron star with a mass of $\sim1.88\,\mathrm{M}_\odot$ (\cite{Quaintrell2003}) and a spin period of ${\sim283 \text{ s}}$ (\cite{Rappaport1975}). Vela X-1 has an orbital period of $\sim8.96$ days (\cite{vanKerkwijk1995a}) and an eccentricity of $\sim0.09$ (\cite{Bildsten1997ApJS}). The source is also found to have an apsidal advance \cite{Falanga2015AnA}. Vela X-1 is located at a distance of $\sim1.96$ kpc as found from the parallax measurements in Gaia Early Data Release 3 \cite{2021AJ....161..147B}. It is known to have strong stellar wind with the mass loss rate of ${\sim10^{-6} \mathrm{M}_{\odot}\text{ yr}^{-1}}$ along with wake structures showing orbital phase dependence in photoelectric absorption (\cite{Nagase1986PASJ},\cite{Tamang2025AnA}).

    \item GX 301-2 is a well-known HMXB system consisting of a highly magnetized neutron star in orbit around the blue hypergiant companion Wray 977 (\cite{white1983accretion}). The system exhibits a relatively long pulse period of approximately 680 seconds and a highly eccentric orbit, with an orbital period of $\sim$41.5 days (\cite{koh1997rapid}). Accretion in GX 301-2 primarily occurs via a dense, inhomogeneous stellar wind, and possibly a transient accretion disk, leading to pronounced variability in the observed X-ray emission, particularly around periastron passage (\cite{leahy2008stellar}). The X-ray spectrum of GX 301-2 is strongly absorbed and often features prominent iron fluorescence lines (\cite{furst2011study}).
\end{itemize}
\subsection{BH-XRBs}
\begin{itemize}
    \item Cyg X-1 was discovered by the UHURU satellite in 1971 (\cite{tananbaum1971}). It is a persistently bright black hole X-ray binary that has been extensively studied. Using VLBA observations (\cite{miller-jones2021}), the mass, distance and inclination of the source were estimated to be $21.2\pm2.2$ M$_{\odot}$, $2.22_{-0.17}^{+0.18}$ kpc (\cite{miller-jones2021}) and 27.51$^{\circ}$, respectively. Cyg X-1 is known for its variable behavior, transitioning between different spectral states. Typically, it remains in a LHS, but occasionally exhibits transitions to HSS. During LHS and IMS, the PDS show multiple broad Lorentzian features along with band limited noise. In soft states of the source, the broad features are absent and PDS show a steep cut-off power law with index of $\sim1$ (\cite{ankur_2021_cygx1}). During transitions from LHS to HSS, the central frequencies of the Lorentzian are observed to shift to higher frequencies (\cite{axelsson2005}). 

    \item Cyg X-3 falls under the category of an HMXB system whose companion star is a Wolf-Rayet star  \cite{vankerkwijk1992}. It is a persistent X-ray source, having an unusually low orbital period of 4.8 hours, which is known for its huge radio outbursts (\cite{parsignault1972}), and luminosity near the maximum accretion power of a solar mass compact star being in the range of L$_{\mathrm{Edd}}$  (\cite{vilhu2009}) for a 10 kpc distance (\cite{bonnetbidaud1988},\cite{tavani2009}). The nature of the compact object in this binary system is still unknown. Recent observations from NASA’s Imaging X-ray Polarimetry Explorer (IXPE; \cite{2016SPIE.9905E..17W}) indicated Cyg X-3 is possibly connected to the first Galactic Ultra-luminous X-ray source (\cite{Veledina2024CygnusX3IXPE}). Its connection with a compact object as well as the Wolf–Rayet star is very rare, making the system a likely progenitor of a double-degenerate system that could eventually produce gravitational waves (\cite{Veledina2024CygnusX3IXPE}).
\end{itemize}

\section{Observation and Data Reduction}
\label{obs_data_red}

Since starting operations in January 2024, XSPECT has observed a wide range of X-ray sources, including NS-LMXBs, BH-XRBs, and Pulsars.
This workshop utilizes XSPECT observations of three NS-LMXBs: Sco X-1 (50 days, 484 ks) Aql X-1 (8 days, 104 ks), and Cir X-1 (17 days, 261 ks); two pulsars: Vela X-1 (28 days, 408 ks), and GX 301-2 (34 days, 380 ks); and two BH-XRBs: Cyg X-1 (91 days, 1210 ks), and Cyg X-3 (16 days, 190 ks). However, only three to four days of data from each source were made available during the workshop, so the participants can focus on learning the data analysis techniques and carry out focused work on a subset of the full available data.
The details of the XSPECT observations utilized in the workshop are presented in Table \ref{Source_observation_log}.
The data utilized during the workshop was chosen to capture certain interesting phenomena observed from the sources such as tracing of the complete Z-track (Sco X-1), presence of thermonuclear X-ray burst (Aql X-1), capturing the dipping as well as high intensity state (Cir X-1), flaring durations (GX 301-2), varying intensity levels (Cyg X-1) and so on. 

As part of data reduction, the Level-1 data is first screened using the nominal filtering criteria recommended by the XSPECT team. The criteria include pointing offset from the source position ($<0.1^{\circ}$), time since SAA exit ($>100$~s), earth avoidance angle ($>5^{\circ}$) etc, and it ensures that data during only the good time intervals (GTIs) are used in further analysis. The screening is done using \texttt{xspl2screen}, which produces the Level-2 (filtered) event files. Subsequently, \texttt{xspevtmerge} is used to merge individual day-wise event files to generate a merged event file. The spectra and light curve for both $2^{\circ} \times 2^{\circ}$ and $3^{\circ} \times 3^{\circ}$ FOV detectors are generated using the \texttt{xspl2specgen} and \texttt{xspl2lcgen} tasks, respectively. The \texttt{xspfilterflare} task further checks the GTI for signs of particle flares and removes the intervals, if any. 

The FOV-wise light curves are added together to create the net source light curve. For timing analysis of pulsars, before generating light curves, barycentric correction is necessary to correct the event arrival times to the solar system barycenter. This is done using the \texttt{xspbary} task, which takes the Level-2 event file and the orbit file (Level-1 product) as input, and produces the barycenter-corrected event file, which is then used for light curve generation. The spectra of the $2^{\circ} \times 2^{\circ}$ and $3^{\circ} \times 3^{\circ}$ FOV detectors are considered independently for spectral fitting, as they have slightly different effective areas due to the different collimators. The \texttt{xspbackspecgen} and \texttt{xspbacklcgen} modules, respectively, take care of the background spectra and light curve generation for further analysis. 

The XSPECT pipeline software which includes the necessary data processing tools (such as data screening and filtering modules, light curve and spectra generation etc), as well as the instrument response (ARF and RMF) and background files generated by the XSPECT team, is hosted at the Indian Space Science Data Center (ISSDC) and is available for download at PRADAN\footnote{\href{https://pradan1.issdc.gov.in/x01}{https://pradan1.issdc.gov.in/x01}}. 

\section{Analysis and Modeling} \label{ana_modeling}

\subsection{Temporal Analysis}

\begin{figure*}
\centering
\includegraphics[width=0.55\textwidth]{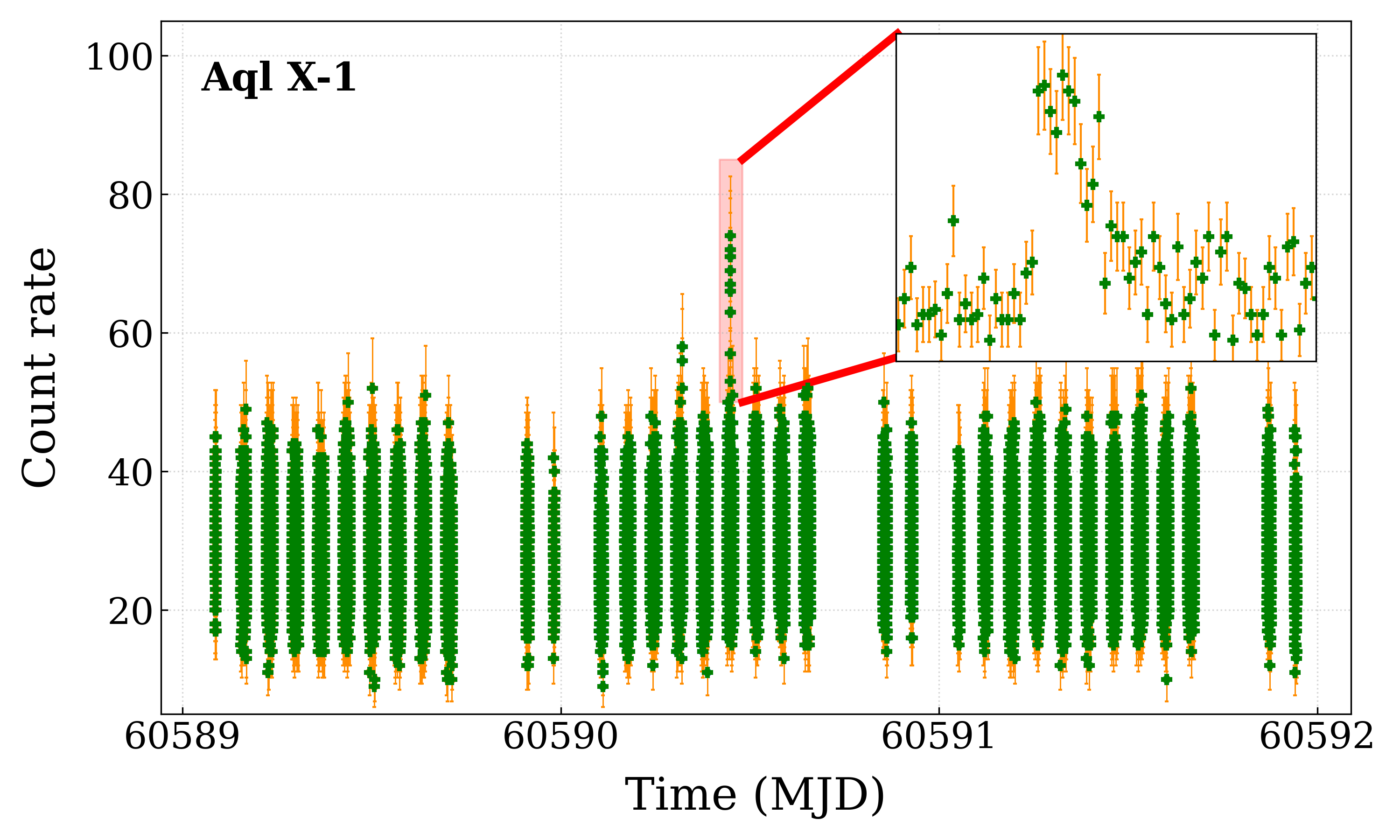}
\includegraphics[width=0.44\textwidth]{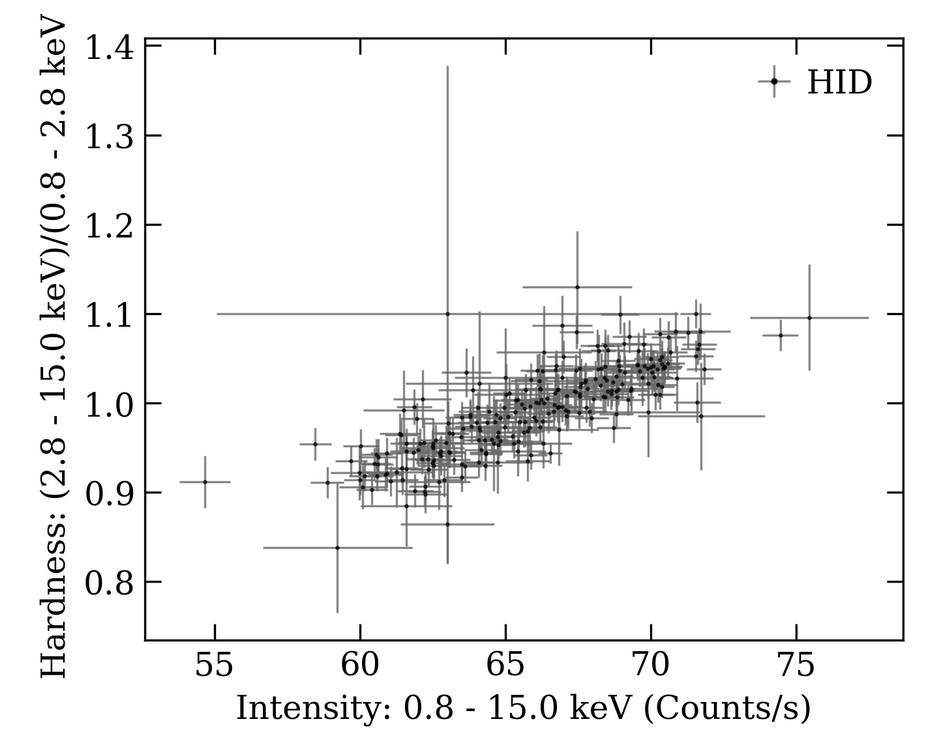}
\caption{(Left) Light curve of Aql X-1 with 1 s binning, obtained in the $0.8–15.0$ keV energy range with the X-ray burst marked. The type-I X-ray burst is shown in the inset. (Right) HID of the source during the observation.}
\label{AqlX-1_Lcurve}
\end{figure*}

\begin{figure*}
    \centering
    \includegraphics[width=0.42\textwidth]{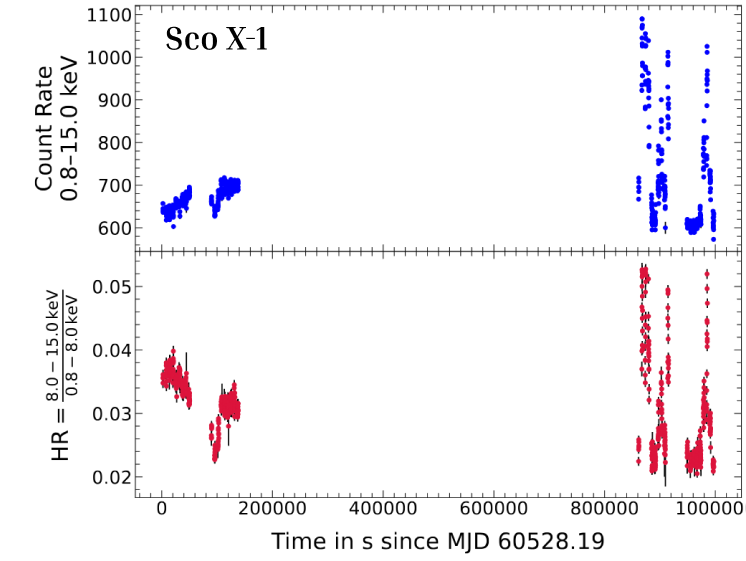}
    \includegraphics[width=0.54\textwidth]{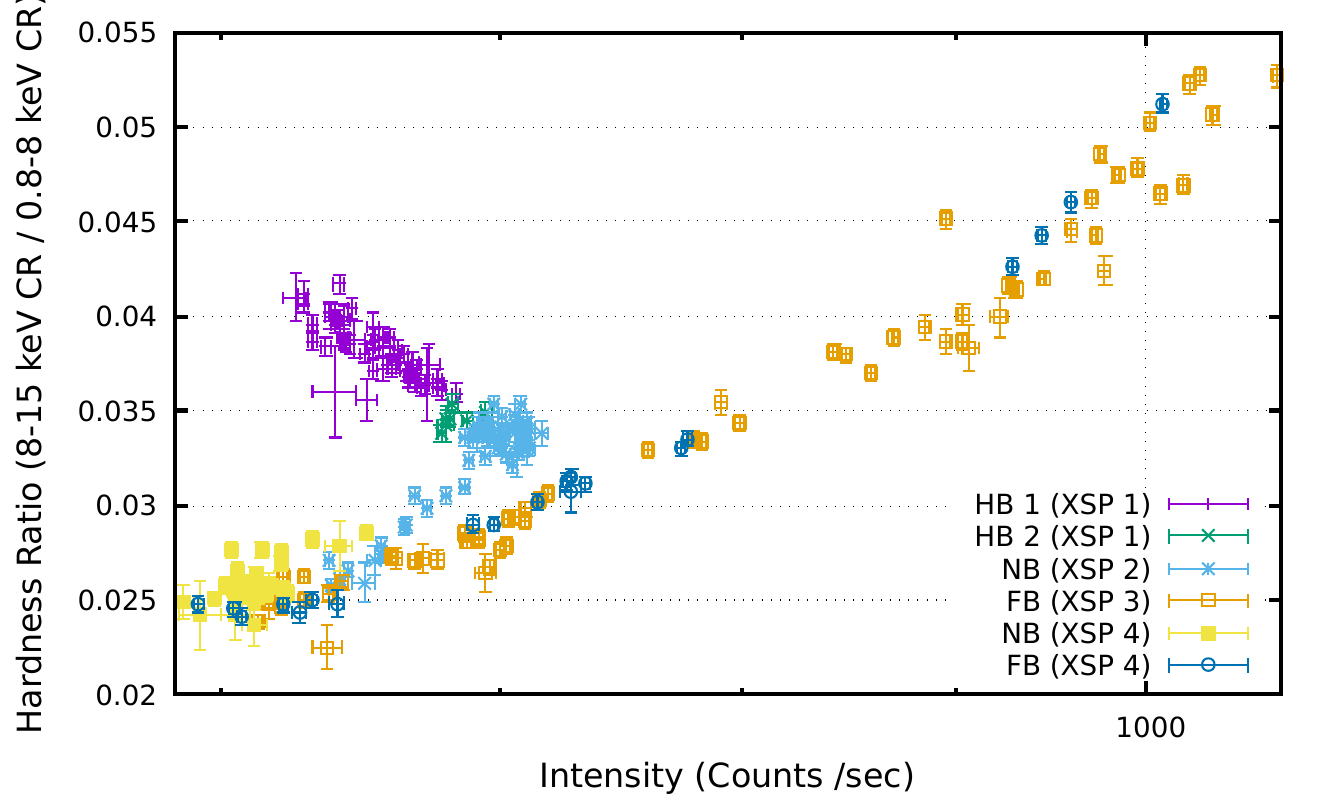}
    \caption{(Left) Sco X-1 light curve with time bin size of 100 s in the 0.8-15.0 keV energy band along with HR. (Right) CCD of Sco X-1 with time bin size of 200 s for $2^{\circ} \times 2^{\circ}$ FOV detectors for each day (denoted as XSP~1, 2, 3, and 4 respectively) of observation. }
    \label{sco_lc_hid}
\end{figure*}

\begin{figure*}
    \centering
    \includegraphics[width=0.5\linewidth]{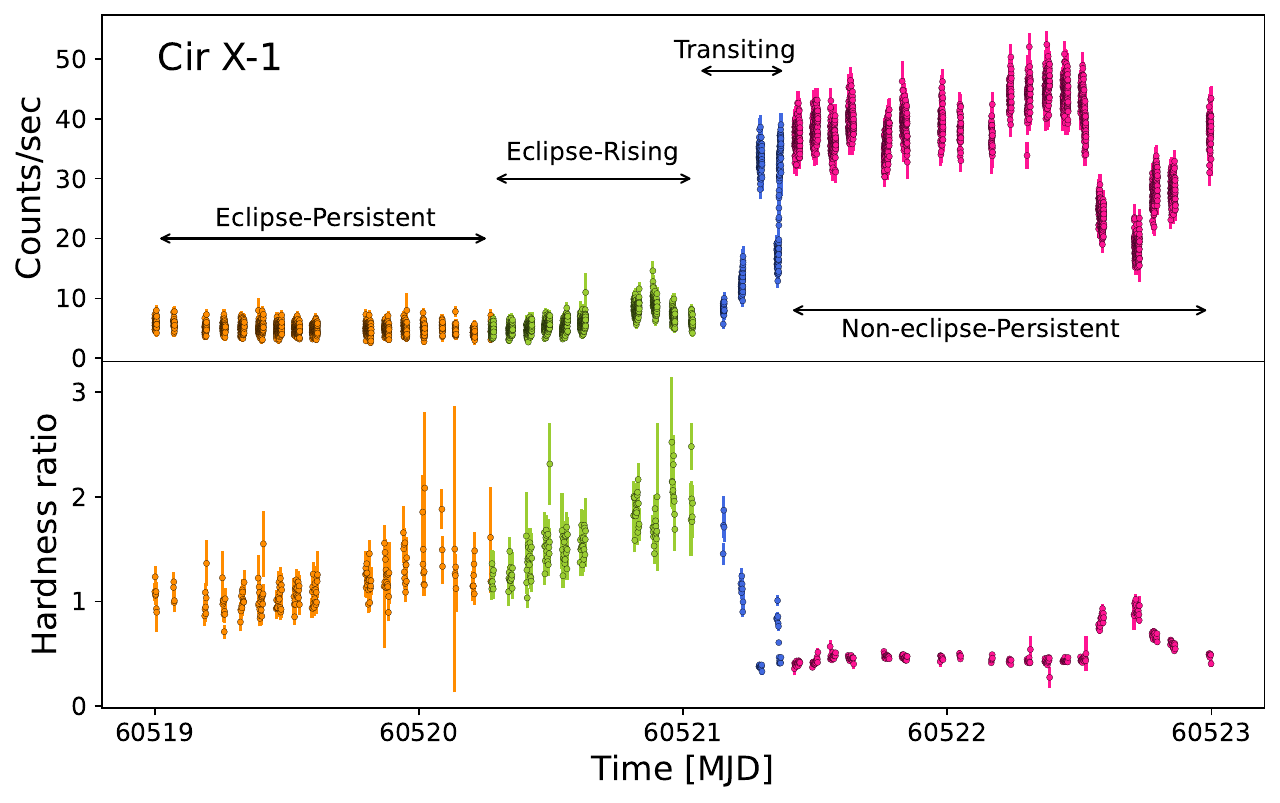}
    \includegraphics[width=0.4\linewidth]{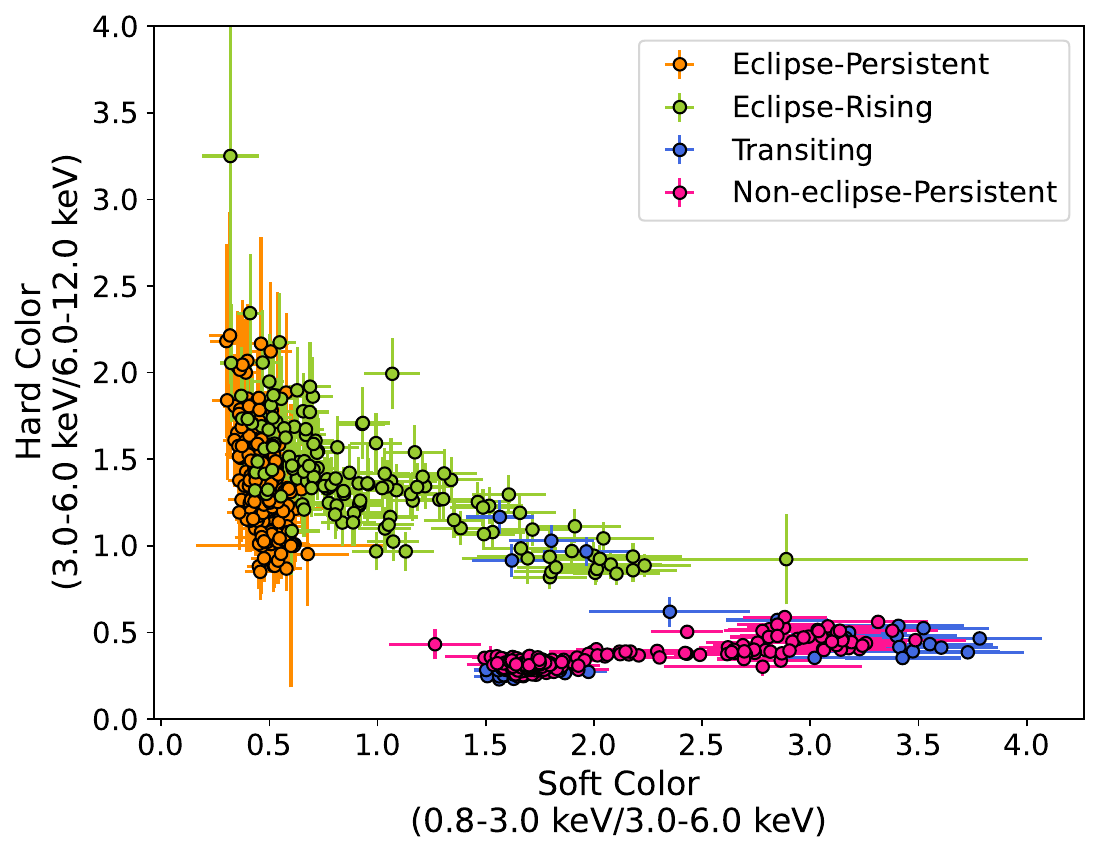}
    \caption{(Left) light curve of Cir X-1 during XSPECT observations. Eclipse and non-eclipse states are marked with horizontal arrow lines along with the HR. (Right) CCD of Cir X-1 for the 4-days of observations is shown.}
    \label{CirX-1_lc_hr}
\end{figure*}

Light curves were extracted with different bin sizes depending on the nature of the source. For Aql X-1, 1 s binned light curve was generated as shown in the left panel of Figure \ref{AqlX-1_Lcurve}. The soft ($0.8-2.8$~keV) and hard ($2.8-15.0$~keV) band light curves were generated to plot the HID (right panel of Figure~\ref{AqlX-1_Lcurve}). Sco X-1 light curves were obtained with 100~s bins in the 
soft ($0.8-8.0$ keV) and total ($8.0-15.0$ keV) bands for FOV of $2^{\circ}\times2^{\circ}$ as displayed in Figure \ref{sco_lc_hid}). Similarly, for Cir X-1, 10 s binned light curves were produced in the 0.8–15.0 keV band, along with corresponding soft ($0.8–5.0$ keV) and hard ($5.0–15.0$ keV) bands to compute the hardness ratio (HR). In addition to HID, a color–color diagram (CCD) was generated for Cir X-1 using soft color ($0.8–3.0$ keV / $3-6$ keV) and hard color ($3–6$~keV / $6-12$~keV) ratios to study the spectral states and variability as shown in Figure \ref{CirX-1_lc_hr}.

\begin{figure*}
    \includegraphics[width=0.47\textwidth]{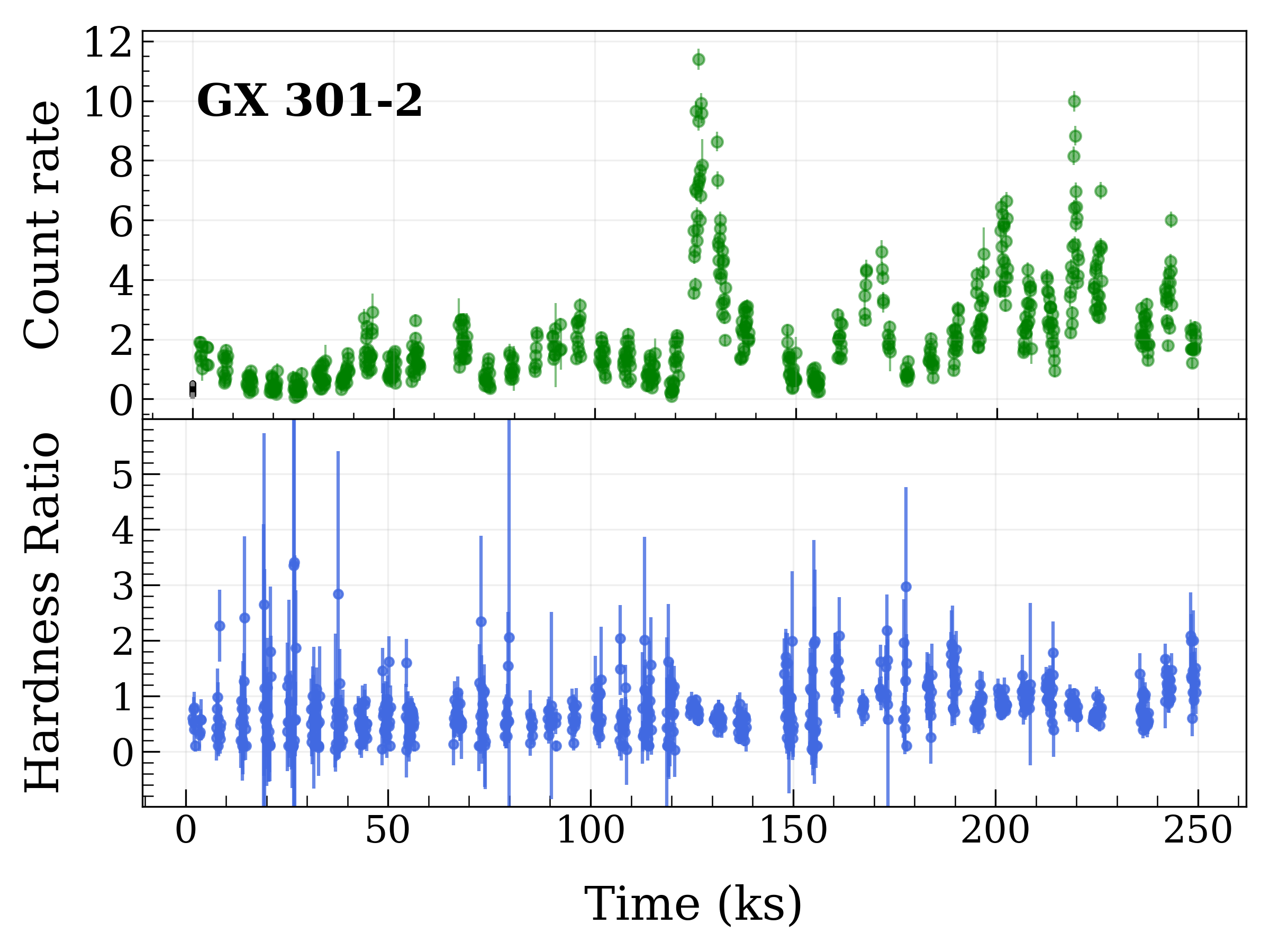}
    \includegraphics[width=0.48\textwidth]{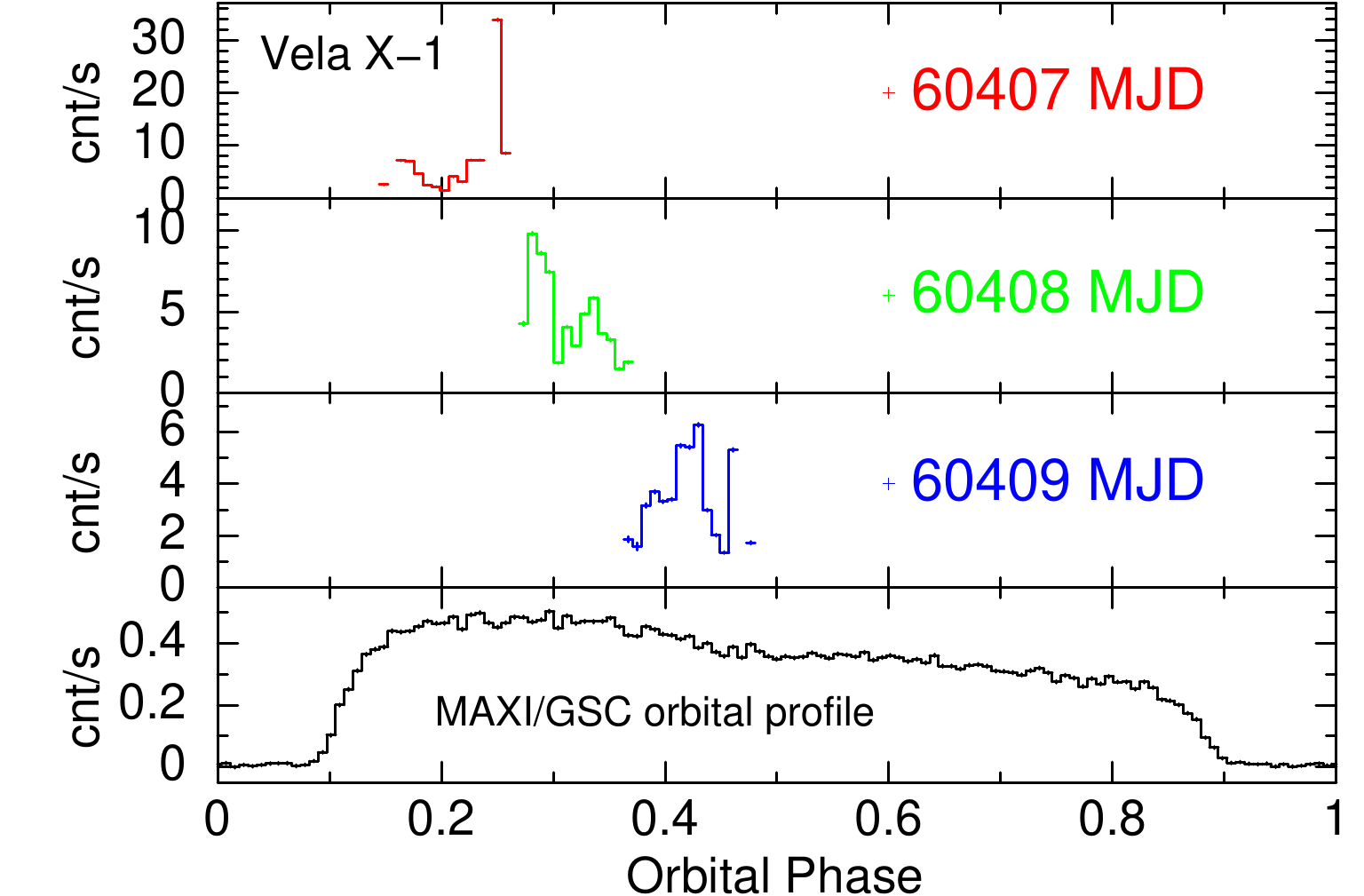}
    \caption{(Left) Background-subtracted light curves in the energy band 0.8-15 keV from FOV $2^{\circ}\times2^{\circ}$ (top panel), along with the corresponding HR (0.8-6.0 keV)/(8-15 keV) in the bottom panel. A flare like feature and gradual hardening with time is visible. (Right) Orbital profile of Vela X-1 with the 0.8$-$8.0 keV XSPECT light curves combined for both FOVs in the top three panels and the bottom panel with the 2$-$20 $\mathrm{keV}$ \textrm{MAXI}/GSC data. }
    \label{GX_vela_lc_hr}
\end{figure*}

For pulsar timing analysis, barycentric corrections were applied during the analysis using \texttt{xspbary} task.
For GX 301-2, 10 s binned light curves were generated to study long-term variability, HID and HR variations were studied to examine spectral changes during the three-day observation as shown in the left panel of Figure \ref{GX_vela_lc_hr}. Right panel of Figure \ref{GX_vela_lc_hr} displays the XSPECT light curve in the 0.8–8.0 keV band for Vela X-1 compared with the 2–20 keV MAXI/GSC orbital profile. 

\begin{figure}
\centering
%FIGURE-1
\includegraphics[width=0.48\textwidth]{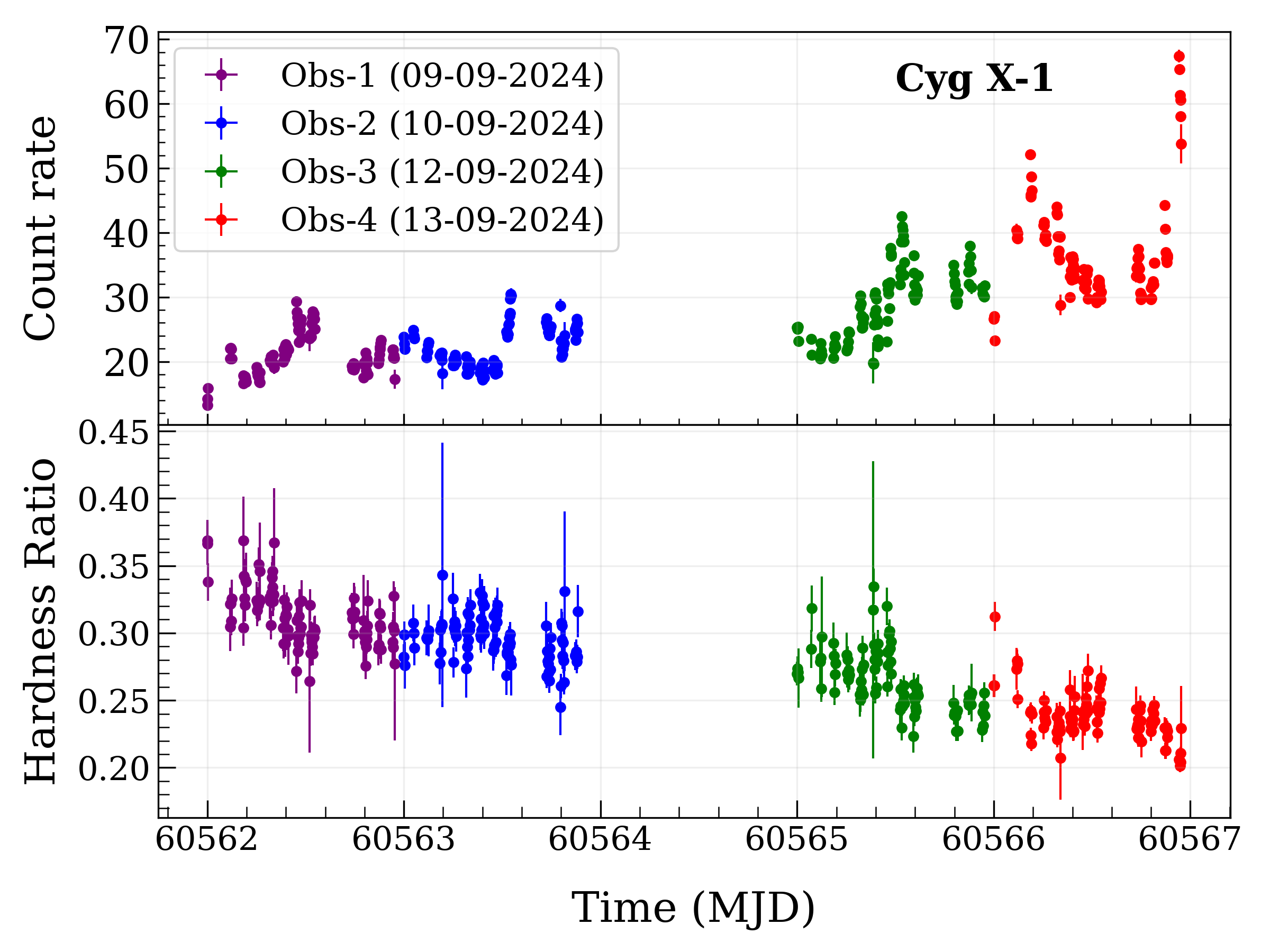}
 \caption{Light curve and HR (defined as the ratio of counts in $4-15$ keV and $0.8-4.0$ keV) of Cyg X-1 from 9$^{\mathrm{th}}$ to 13$^{\mathrm{th}}$ September 2024. The different colors corresponds to different day which is shown in the legend in lower panel. See text for details.}
 \label{CygX-1_lc_hid}
\end{figure}

\begin{figure}
    \centering
    \includegraphics[width=0.45\textwidth]{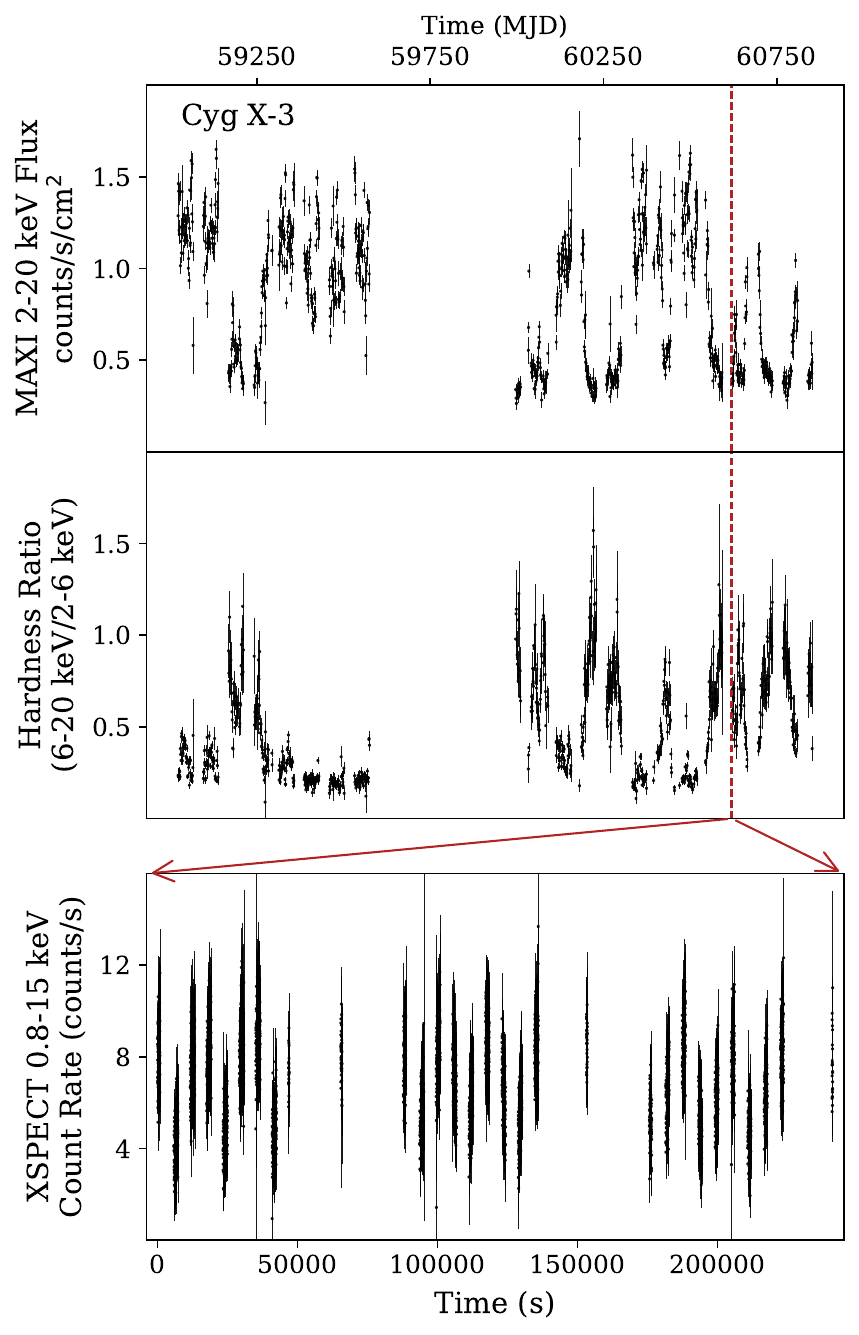}
    \caption{Top panel shows the long-term variations in the count rate of Cyg X-3, as represented by the 1-day binned light curve in the 2-20 keV range from MAXI/GSC. Middle panel displays the corresponding HR for the MAXI/GSC observations in the energy bands of $6-20$ keV and $2-6$ keV. The red dashed line in both panels indicates the timing of the XSPECT observation we have considered. Bottom panel shows the 10-second binned background subtracted light curve in the $0.8-15$ keV range from the total XSPECT observation.}
    \label{CygX-3_lc_hr}
\end{figure}

In the case of BH sources, the light curve and HR are plotted to understand the variability in the sources.
For Cyg X-1, 200 s binned light curves were generated in the $0.8–15$ keV band using $2^{\circ}\times2^{\circ}$ FOV for the observations on 9$^{\mathrm{th}}$, 10$^{\mathrm{th}}$, 12$^{\mathrm{th}}$, and 13$^{\mathrm{th}}$ September 2025, and the HR was calculated as the count rate in $4–15$ keV to that in $0.8–4$ keV (see Figure \ref{CygX-1_lc_hid}). For Cyg X-3, the MAXI $2–20$ keV light curve and HR ($2–6$ keV / $6–20$ keV) were used to study its long-term behavior, showing the source in a soft state before transitioning to a harder state. This is shown in Figure \ref{CygX-3_lc_hr} along with the XSPECT observations.

In order to study the variability in shorter scale, we generated 0.02 s binned lightcurve in the energy range $0.8-15$ keV considering only FOV $2^{\circ}\times2^{\circ}$ detectors. 
We produced Poisson noise subtracted PDS using \texttt{powspec} considering 8192 bins per segment and Nyquist frequency of 25 Hz.

\subsection{Spectral Analysis}
The spectral analysis is carried out with the help of \texttt{XSPEC} with $\chi^{2}$ statistics, quoting uncertainties at the 1$\sigma$ level unless specified otherwise. For Aql X-1, an energy range of $0.8-10$ keV was considered for analysis.
A single absorbed component could not describe the persistent emission, so a composite model \texttt{edge*tbabs(Nthcomp + Gaussian + diskbb)} was adopted, where \texttt{tbabs} accounts for interstellar absorption, \texttt{diskbb} represents multicolor disk emission, and \texttt{Nthcomp} models thermal Comptonization. A narrow Gaussian line near 6.7 keV fits Fe K$\alpha$ emission.
A Type I thermonuclear burst showing a fast-rise exponential-decay (FRED) profile ($\sim$5 s rise, $\sim$32 s decay time) was also detected, dominant in the 4–8 keV band. Time-integrated burst spectra were fitted with both double \texttt{bbodyrad} and a more physical neutron-star atmosphere model \texttt{tbabs*const(nsatmos + bbodyrad + powerlaw)}.
For Sco X-1, spectra were analyzed branch-wise in the 0.8-15.0 keV along the Z-track by segmenting the HID to trace spectral evolution. The initial model \texttt{const*tbabs(ThComp*bbodyrad + Gaussian)} assumed seed photons from the neutron-star surface, but the inferred blackbody radius ($\sim$248 km for 2.8 kpc) was un-physical. Replacing \texttt{bbodyrad} with \texttt{diskbb} yielded the improved model \texttt{const*tbabs*(ThComp*diskbb + Gaussian)}, attributing seed photons to the accretion disk. Two edge components at $\sim1.5$~keV and $\sim1.8$~keV were included to account for the K-absorption edges of Al and Si. The covering fraction of \texttt{Thcomp} was fixed at unity for consistency, and fluxes were derived using \texttt{cflux} to estimate luminosities in Eddington units.
For Cir X-1, spectral fitting in the $0.8–8.0$ keV range was performed in different orbital phases. 
We performed a detailed X–ray spectral analysis of the source across four distinct phases such as Eclipse-Persistent (EP), Eclipse-Rising (ER), Transiting (Tr), and Non-eclipse-Persistent (NeP) using the model \texttt{const * tbabs * (pcfabs*(diskbb+ Nthcomp) + Gaussian)}. This composite model captures both thermal emission from the accretion disk and the Comptonized hard X–ray tail, while also allowing for intrinsic and interstellar absorption. The inclusion of a Gaussian line component accounts for the Fe $K_{\alpha}$ emission.

The spectra of both GX 301-2 and Vela X-1 were modeled using phenomenological approaches. For GX 301, absorption was accounted for using the \texttt{tbabs} and \texttt{pcfabs} models, while the continuum emission was described by a simple power-law with two \texttt{Gaussian} components added to represent the fluorescent iron lines near 6.4 keV and 7.0 keV. 
For Vela X-1, both phase-averaged and time-resolved spectra were modeled using a partially absorbed Comptonized continuum along with a fluorescent iron emission line, described by the model \texttt{const*tbabs*pcfabs*(compST + Gaussian)}. An alternative model employing a blackbody for the continuum yielded poor fits with unconstrained parameters and unphysically high temperatures.
The same model was used for the time-resolved analysis, with the \texttt{Gaussian} width parameter ($\sigma_{\mathrm{Fe}}$) kept fixed.

For Cyg X-1, we fitted the spectra using the model combination \texttt{const*tbabs(Nthcomp + diskbb + Gaussian)}. The \texttt{diskbb} model (\cite{mitsuda1984_diskbb},\cite{makishima1986_diskbb}) describes the thermal emission from the accretion disk, and \texttt{Nthcomp} (\cite{zdziarski1996_Nthcomp}) represents the thermal Comptonization of seed photons from the disk that are reprocessed in the corona, with the seed photon temperature ($kT_{\mathrm{bb}}$) tied to the inner disk temperature ($kT_{\mathrm{in}}$) of \texttt{diskbb}. To address weak residuals near 1.5 keV and 1.8 keV, likely due to Al and Si present in the detector, we included two additional \texttt{edge} components in the model. 
For Cyg X-3, the combined spectrum from three days of observations was analyzed in the $1-9$ keV band, excluding background-dominated data outside this range. 
The spectrum was modeled using phenomenological model components, \texttt{tbabs*pcfabs(diskbb + 4*Gaussian)}, where partial absorption (\texttt{pcfabs}) arises from the clumpy wind of the companion star. Four prominent emission lines corresponding to Si, S XVI$_1$, S XVI$_2$, and Fe at 1.89 keV, 3.09 keV, 2.5 keV, and 6.61 keV, respectively, were modeled using four \texttt{Gaussian} components.

\section{Results and Discussion}\label{red_and_disc}

\subsection{Aql X-1}
XSPECT observed Aql X--1 during the initial decay phase of its 2024 outburst. The X-ray lightcurve of Aql X--1 (right panel of Figure~\ref{AqlX-1_Lcurve}) indicates a relatively stable persistent flux level during the observation with a single thermonuclear burst superimposed. The HID is plotted in the right panel of Figure~\ref{AqlX-1_Lcurve}. The thermonuclear burst exhibits a classical FRED morphology, with burst intensity peaking in the $4–8$ keV range, typical for Type-I bursts in Atoll sources. 
PDS show no evidence of QPOs or significant breaks in the $0.01–15$ Hz frequency band, indicating the variability is dominated by white noise during this observation window.

\begin{figure}
  \centering
  \includegraphics[width=0.48\textwidth]{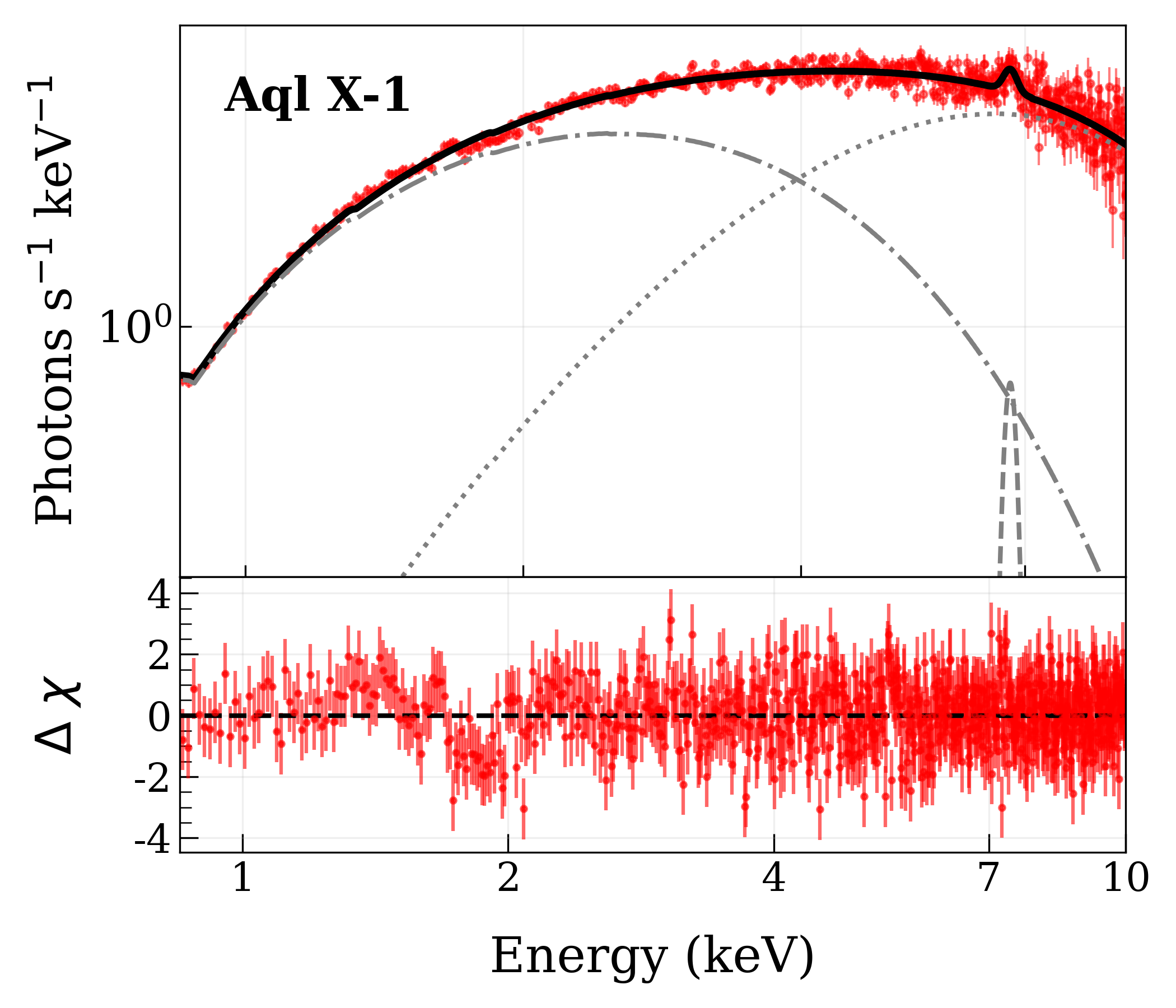}
  \caption{Spectral fitting of Aql X-1 with the model \texttt{tbabs*(diskbb+Nthcomp+Gaussian)} (top panel) and residuals (bottom panel).}
  \label{AqlX1_Spectral_fit}
\end{figure}

The broadband spectrum was well described by a combination of a disk-blackbody, Comptonized corona, and iron line represented as \texttt{edge*tbabs(Nthcomp + Gaussian + diskbb)} as shown in Figure \ref{AqlX1_Spectral_fit}. Spectral fit supports a hot corona with electron temperature $kT_e \sim1.62$~keV and a truncated inner disk extending to $\sim17.2$~km from the neutron star surface. However, the photon index ($\Gamma$) is $\sim 1.07$, indicating a weak Comptonization process. The presence of a narrow Gaussian line at 6.7~keV suggests weak Fe~K$\alpha$ fluorescence originating from reflection off the inner disk.
The time‐integrated fits using a realistic neutron‐star atmosphere model (\texttt{nsatmos}) yielded a stellar radius of $9.94\pm1.56$~km assuming a distance of 4.1–5.9~kpc, broadly consistent with theoretical ranges for canonical neutron stars. A detailed spectro-temporal analysis including the results from time-resolved analysis is presented in \cite{chatterjee2026}.

\subsection{Sco X-1}
The source traces the complete Z-track which include the Horizontal branch (HB), Normal branch (NB) and the Flaring branch (FB) during XSPECT observations as shown in the right panel of Figure \ref{sco_lc_hid}.
In the first two days of observations, the source was found to trace the HB and NB, whereas, during the next two days, the source traced the FB.
PDS suggest a low rms amplitude, which indicate the absence of QPO features in any branch of the source during these observations.

\begin{figure}
\centering
  \includegraphics[width=0.48\textwidth]{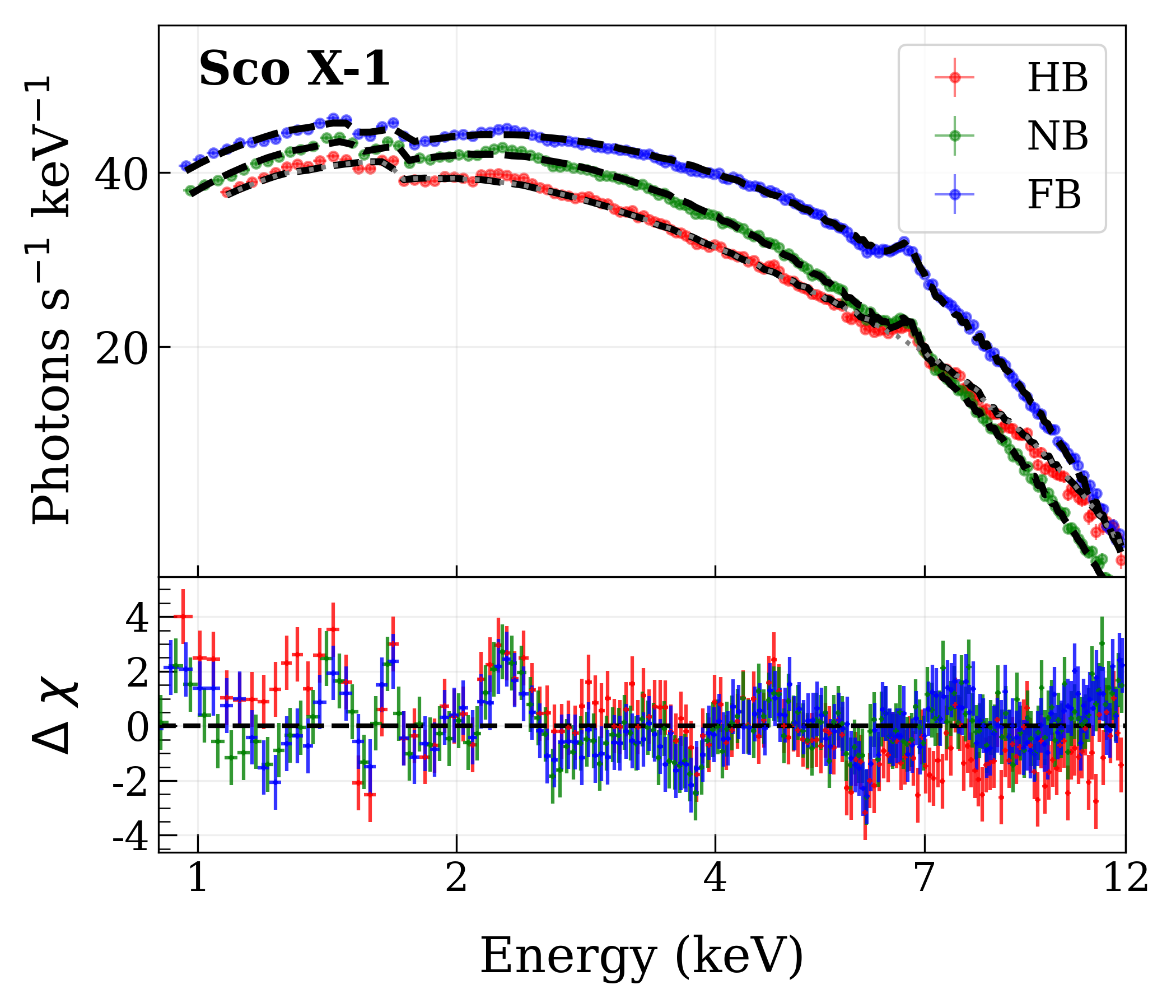}
  \caption{Fitted spectra of Sco X-1 using the best fitted model combination \texttt{const*edge*tbabs*(thcomp*diskbb + Gaussian)} for HB, NB and FB segments.}
  \label{ScoX1_Spectral_fit}
\end{figure}

The spectral fit to the HB, NB and FB spectra are shown in Figure \ref{ScoX1_Spectral_fit}.
The spectral results indicates an evolution of the spectral parameters along the Z-track.
The photon index ($\Gamma$) takes into account the changes in the spectral shape, increasing from 1.7 in HB to 1.83 in NB.
This spectral softening is indicative of a cooling corona, typically associated with an enhanced soft photon flux due to an increased accretion rate. As expected, the temperature of the disk component varies between 0.7 keV and 0.8 keV, showing a modest but consistent increase.
It is to be noted that the disk temperature again cools down $0.6-0.7$ keV as the system moves to FB. More notably, the change in the inner disk radius, estimated from the disk normalization suggests towards a geometric adjustment in the inner accretion flow, potentially caused by increased radiation pressure or local structural changes during the FB.  Additionally, a persistent Fe K$\alpha$ emission line is detected at $\sim$ 6.68 keV across all segments, with a moderate line width of approximately 0.21 keV. The normalization of the \texttt{Gaussian} component shows an anti-correlation with respect to the electron temperature.
This may indicate towards a reduced reflection component in the system and probably indicates a higher ionization in soft state as seen in BH systems.
Further details on the spectral evolution of the source along the various branches of the Z-track are presented in \cite{2025MNRAS.543.3754P}.

\subsection{Cir X-1}

The light curve (see Figure \ref{CirX-1_lc_hr}) shows the transition of the source through four distinct phases, ie. EP, ER, Tr, and NeP. In the EP phase, the source is faint, steady, and spectrally hard, located on the upper-left branch of the CCD, indicating that the central emission is obscured and the observed X-rays likely arise from scattered or coronal emission. During the ER phase, the brightness increases while the hardness remains high, showing the source is still in a hard state. The Tr phase marks a rapid spike in brightness and a sharp drop in hardness, signifying the hard-to-soft state transition driven by the inward movement of the accretion disk and intensified thermal emission. In the NeP phase, the source reaches peak brightness with low hardness, settling into the soft state on the lower-right CCD branch, dominated by un-obscured disk emission. Occasional dips during this phase likely result from temporary obscuration by a thickened outer disk or short-term fluctuations in the accretion flow.

\begin{figure}
\centering
  \includegraphics[width=0.48\textwidth]{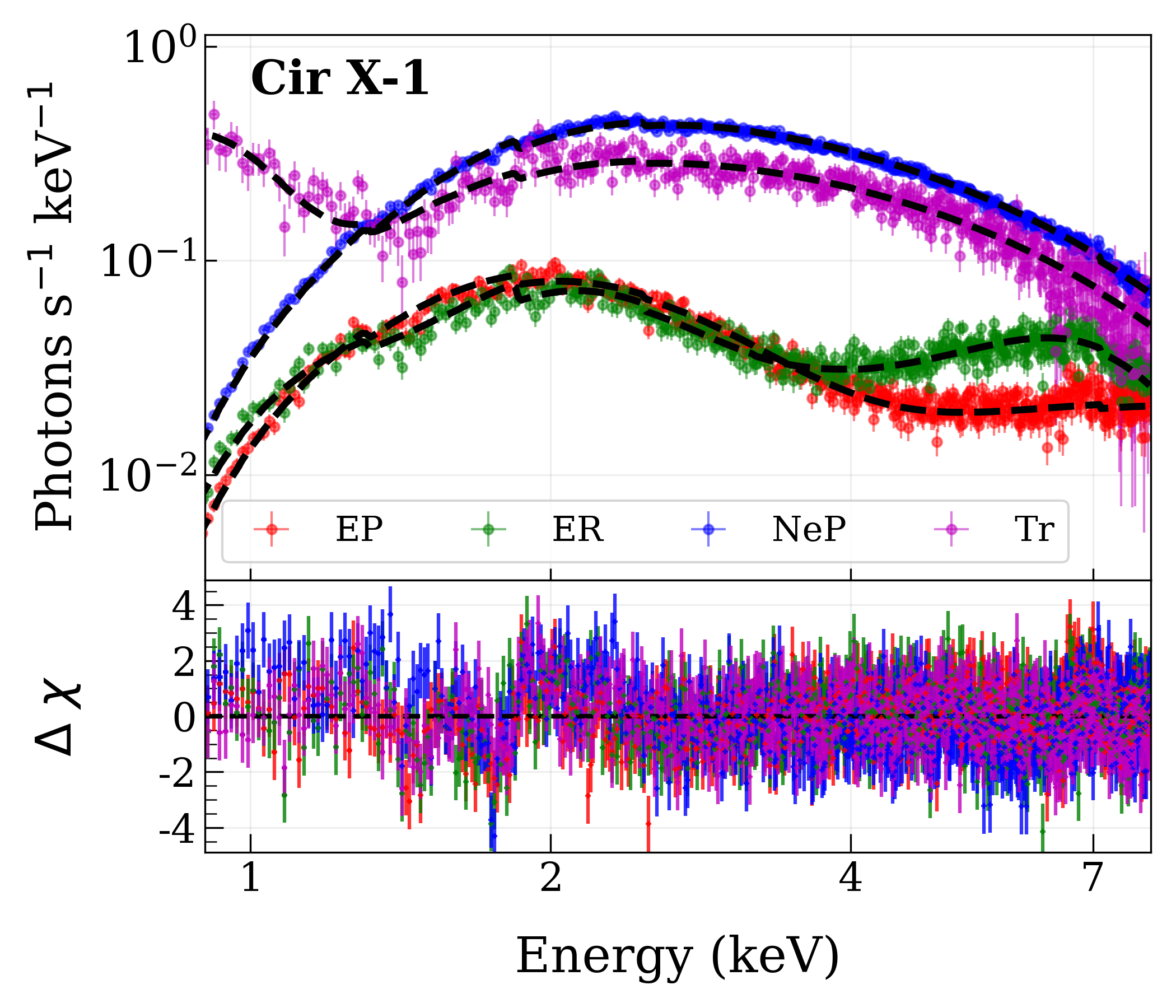}
  \caption{Model fitted X-ray spectra of the 4 phases (EP, ER, Tr and NeP) of observations of Cir X-1 from XSPECT. See text for details.}
  \label{CirX1_Spectral_fit}
\end{figure}

The spectral analysis of Cir X-1 across four orbital phases (EP, ER, Tr and NeP) reveals clear phase-dependent variations in absorption, continuum shape, and line features. 
Spectral fit to these phases are shown in Figure \ref{CirX1_Spectral_fit}.
The interstellar absorption column density, represented by \texttt{tbabs}, was kept fixed and consistent across all phases.
The intrinsic absorber implemented through \texttt{pcfabs} revealed clear variability in both column density and covering fraction.
The ER phase showed the highest level of intrinsic absorption, with $N_{H} \sim 5.17\times10^{22}$ cm$^{-2}$ and a covering fraction close to unity, suggesting the presence of dense, clumpy material partially obscuring the central region. This strong absorption is indicative of structural changes in the inner disk or enhanced outflow activity.
Systematic variations in $T_{in}$ from $0.09-0.56$~keV suggest changes in the inner-disk radius or in the irradiation experienced by the disk. The Comptonization component displayed pronounced phase dependence. The photon index $\Gamma$ increased gradually from EP to NeP, indicating a progressive softening of the Comptonized spectrum.
This trend typically reflects reduced optical depth or enhanced cooling of the hot electron corona. The electron temperature $kT_e$ exhibited a strong increase during the ER phase, reaching nearly 8.5 keV, before dropping again in later phases. This behaviour suggests that the corona experienced a temporary heating episode, possibly linked to enhanced accretion or magnetic activity.
A broad Fe $K_{\alpha}$ line was detected prominently in the ER phase at around 6.63 keV. Its presence only during this phase supports the idea that the geometry of the disk–corona system undergoes significant reconfiguration, allowing fluorescent reprocessing in the disk to become observable. The absence of the line in the other phases may point to either weaker illumination of the disk or stronger obscuration. 
The bolometric Eddington ratio increased systematically from EP to NeP, moving from 0.04 to 0.29. This demonstrates that the system gradually transitioned toward higher luminosity states, consistent with spectral softening and enhanced Compton cooling. 

\subsection{GX 301-2}
\begin{figure*}
    \centering
    \includegraphics[width=0.45\linewidth]{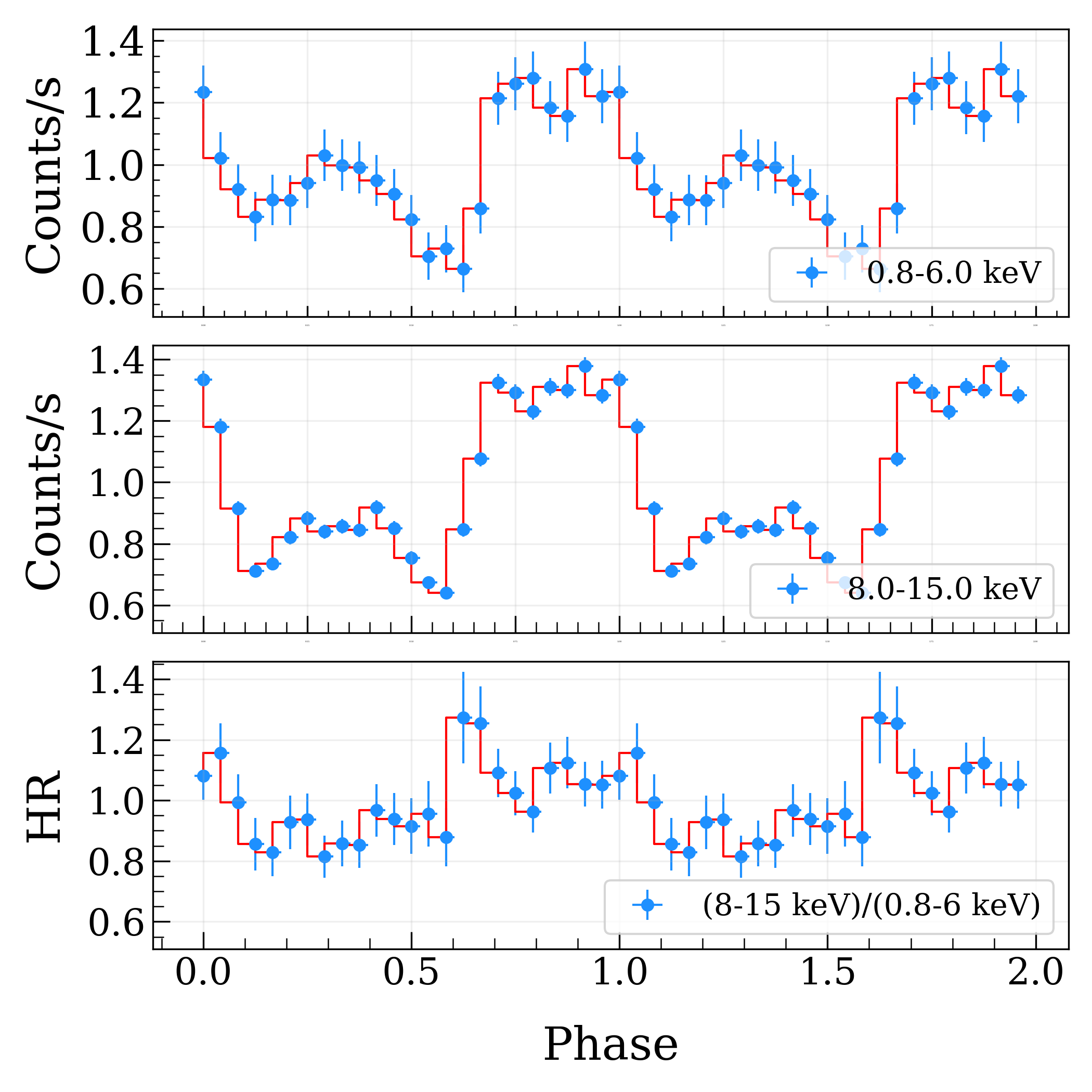}
    \includegraphics[width=0.5\textwidth]{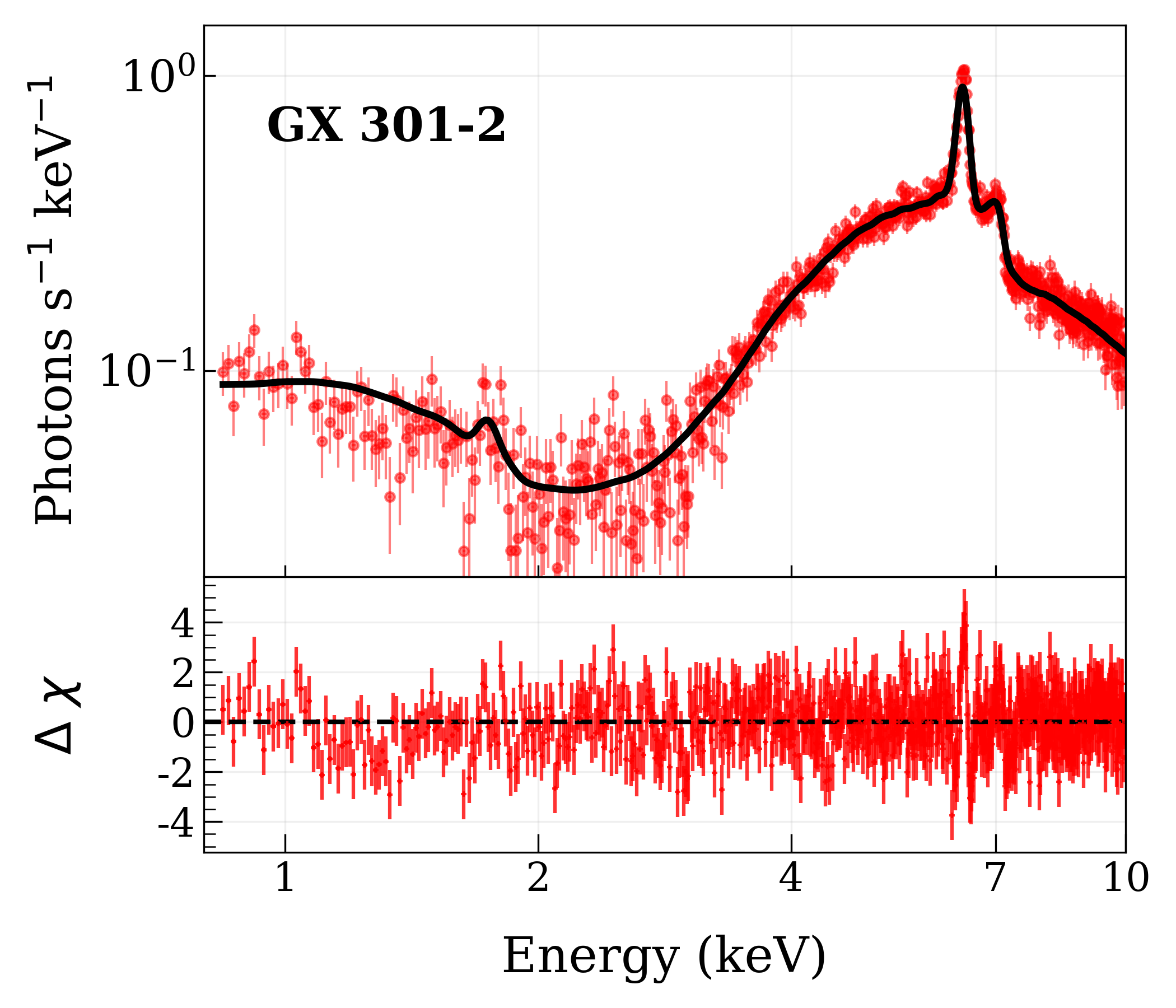}
    \caption{(Left) Folded pulse profiles of GX 301-2 in different energy bands (top to bottom: 0.8–6 keV, 8–15 keV and hardness ratio), showing a double-peaked structure and phase-dependent variation in pulse shape. (Right) The best-fit phase average 0.9$-$10 $\mathrm{keV}$ spectrum of GX 301-2 using the spectral model {\texttt{const*tbabs*pcfabs*(powerlaw + Gaussian + Gaussian)}} for the three days XSPECT data. Bottom panel shows the residuals of the best-fit spectral model. }
    \label{GX301_pulse_profile_spec}
\end{figure*}

During the three-day XSPECT observation, GX 301-2 exhibited significant variability in its X-ray emission and HR, as shown in the light curve (left panel of Figure ~\ref{GX_vela_lc_hr}). 
A clear spectral hardening was observed during periods of increased intensity, suggesting enhanced absorption or changes in the emission properties of the source during flaring episodes. In HID, at higher intensities (above 2 cts s$^{-1}$), the hardness ratio remained nearly constant around 0.5, implying that intrinsic emission dominated over absorption during flares. The spin period of GX 301-2, determined through epoch-folding, was found to be $671.4 \pm 0.1$ seconds. Energy-resolved pulse profiles, folded with this period, revealed a clear double-peaked structure, with subtle energy dependence. The pulse-profile of the source in different energy bands are shown in the left panel of Figure \ref{GX301_pulse_profile_spec}.

\begin{figure}
    \centering
    \includegraphics[width=0.48\textwidth]{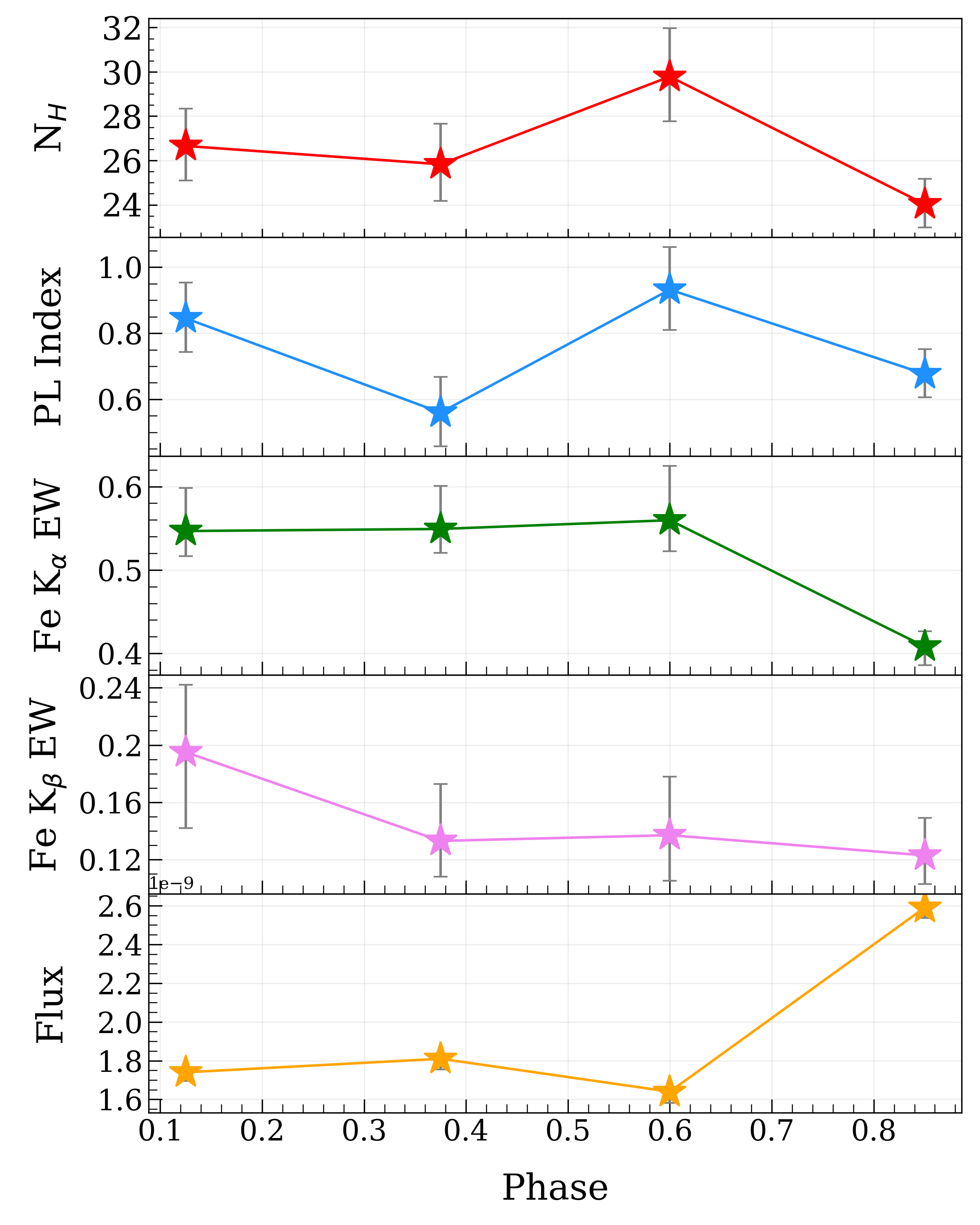}
    \caption{ Phase resolved spectral parameter variation of hydrogen column density, power law index of the emission, equivalent width (keV) of two Gaussian lines, and flux for GX 301-2 from the \texttt{tbabs*(Nthcomp+Gaussian+Gaussian)} model.}
    \label{phase_reso_GX301}
\end{figure}

The time-averaged spectrum of GX 301-2 in the 0.8–10 keV range was well-fit by a model consisting of an absorbed power-law continuum, and \texttt{Gaussian} components for Fe K$\alpha$ and K$\beta$ emission lines, along with partial covering absorption. The fit captured the overall spectral shape, with strong absorption at lower energies and prominent iron line features around 6–7 keV. Time-resolved spectral analysis over five intervals revealed significant variability. In the lower flux intervals, the spectra were softer, while during brighter intervals, the spectra became harder with stronger iron line features. These trends supported the hardness-intensity relation: brighter states were associated with harder emission. Variations in the hydrogen column density ($N_{\rm H}$), which fluctuated between $2\text{–}3 \times 10^{23}$ cm$^{-2}$, were also observed, likely due to clumpy structures in the stellar wind. The equivalent width of the Fe K$\alpha$ line increased in intervals with higher $N_{\rm H}$, indicating enhanced reprocessing. Additionally, phase-resolved spectral analysis was also performed, which revealed systematic variations in the spectral parameters with the neutron star’s spin period as shown in Figure \ref{phase_reso_GX301}. The column density modulated with the pulse phase, peaking around phase 0.5–0.7, suggesting periodic obscuration of the emission region by accretion material. The Fe K$\alpha$ equivalent width showed a pronounced dip during the pulse phase 0.7–1.0, while the Fe K$\beta$ equivalent width remained relatively constant, especially during flux maxima.

\subsection{Vela X -1}

\begin{figure}
    \centering
    \includegraphics[width=0.48\textwidth]{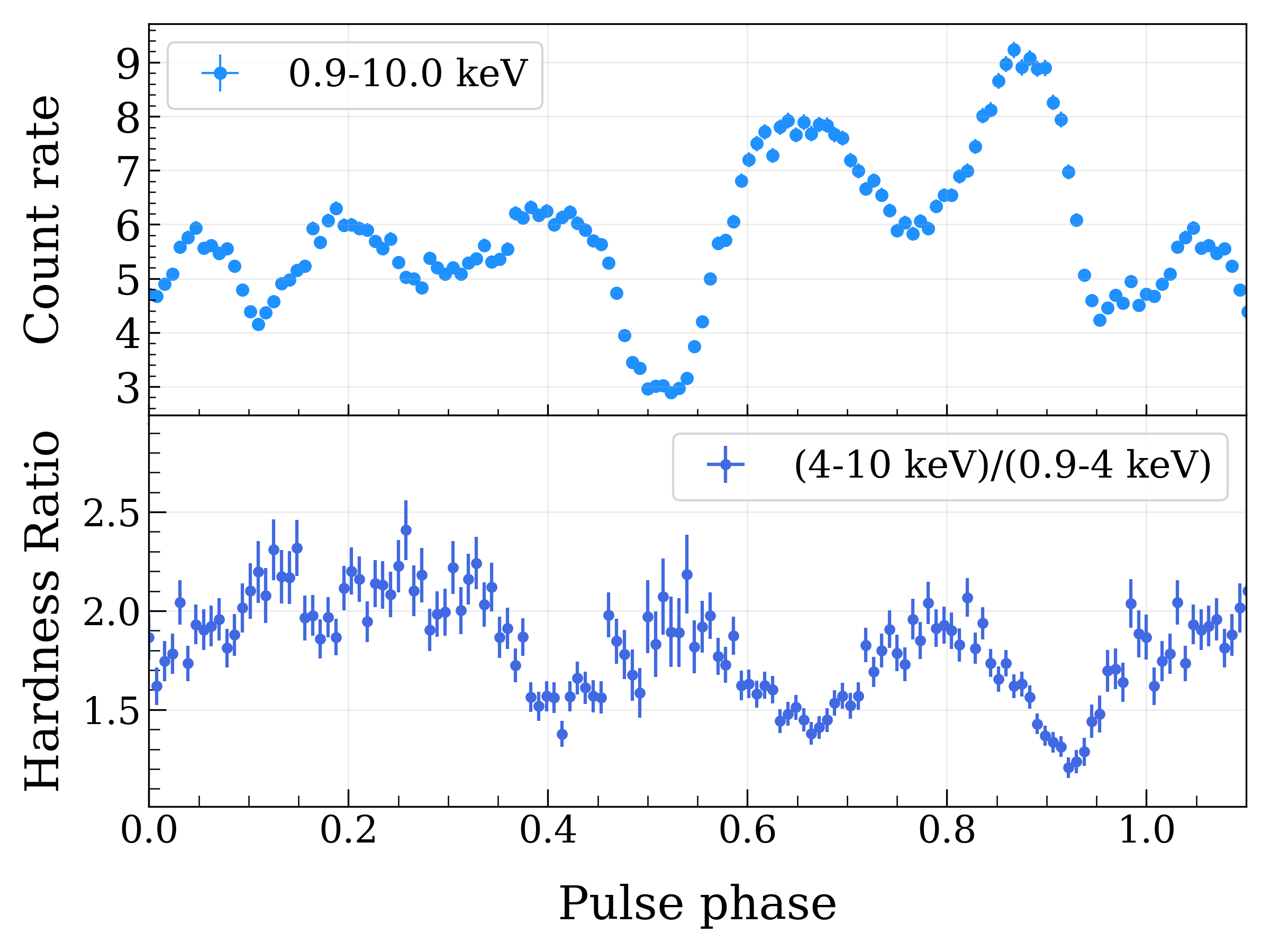}
     \includegraphics[width=0.46\textwidth]{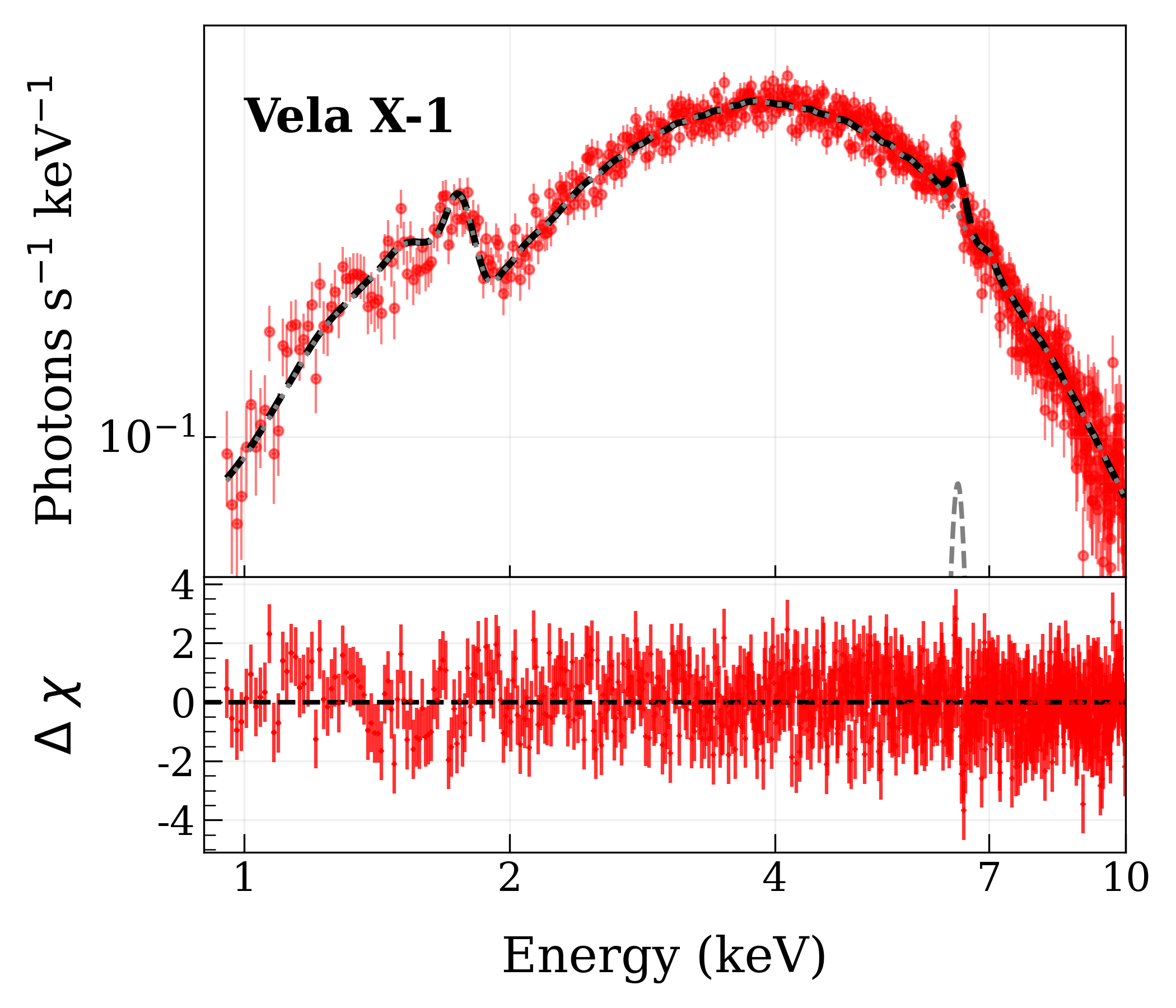}
    \caption{(Top) The pulse profile of Vela X-1 with the spin period, $P_s=283.547$, in the energy ranges, 0.9$-$10 keV (panel 1).  Variation in hardness ratio (blue), $HR=H/S$, with the spin phase as calculated from the energy bands, $S$: 0.9$-$4 $\mathrm{keV}$ and $H$: 4.0$-$10 $\mathrm{keV}$ (panel 2). (Bottom) The best-fit phase average 0.9$-$10 $\mathrm{keV}$ spectrum of Vela X-1 using the spectral model {\texttt{const*tbabs*pcfabs*(compST + Gaussian)}} for the three days XSPECT data (panel 1). Residuals of the best-fit spectral model (panel 2).}
    \label{VelaX-1_pulse_profile_spec}
\end{figure}

Orbital profile of the source in the $0.8-8.0$ keV during the XSPECT observations are shown in the right panel of Figure \ref{GX_vela_lc_hr}.
We searched for the spin period using \texttt{efsearch} for the orbital corrected lightcurve in the energy range $0.9-10.0$ keV and found
the best spin period, P$_{\mathrm{s}}$ = 283.547 $\pm$ 0.001.
The pulse profile of the source is shown in the top panel of Figure \ref{VelaX-1_pulse_profile_spec}.

The spectral fit to the phase averaged spectra is shown in the bottom panel of Figure \ref{VelaX-1_pulse_profile_spec}.
The best-fit spectral parameters for the spectral model used are given in Table \ref{table_best_fit} 
We see that the best-fit value of $n_{H}$ rises from the first day to the last day of available data, $\phi_\text{orb}\sim0.15-0.48$ as seen in Table \ref{table_best_fit}. Also, the electron temperature for the comptonization model is almost constant over this orbital phase.

\begin{figure}
    \centering
    \includegraphics[width=0.48\textwidth]{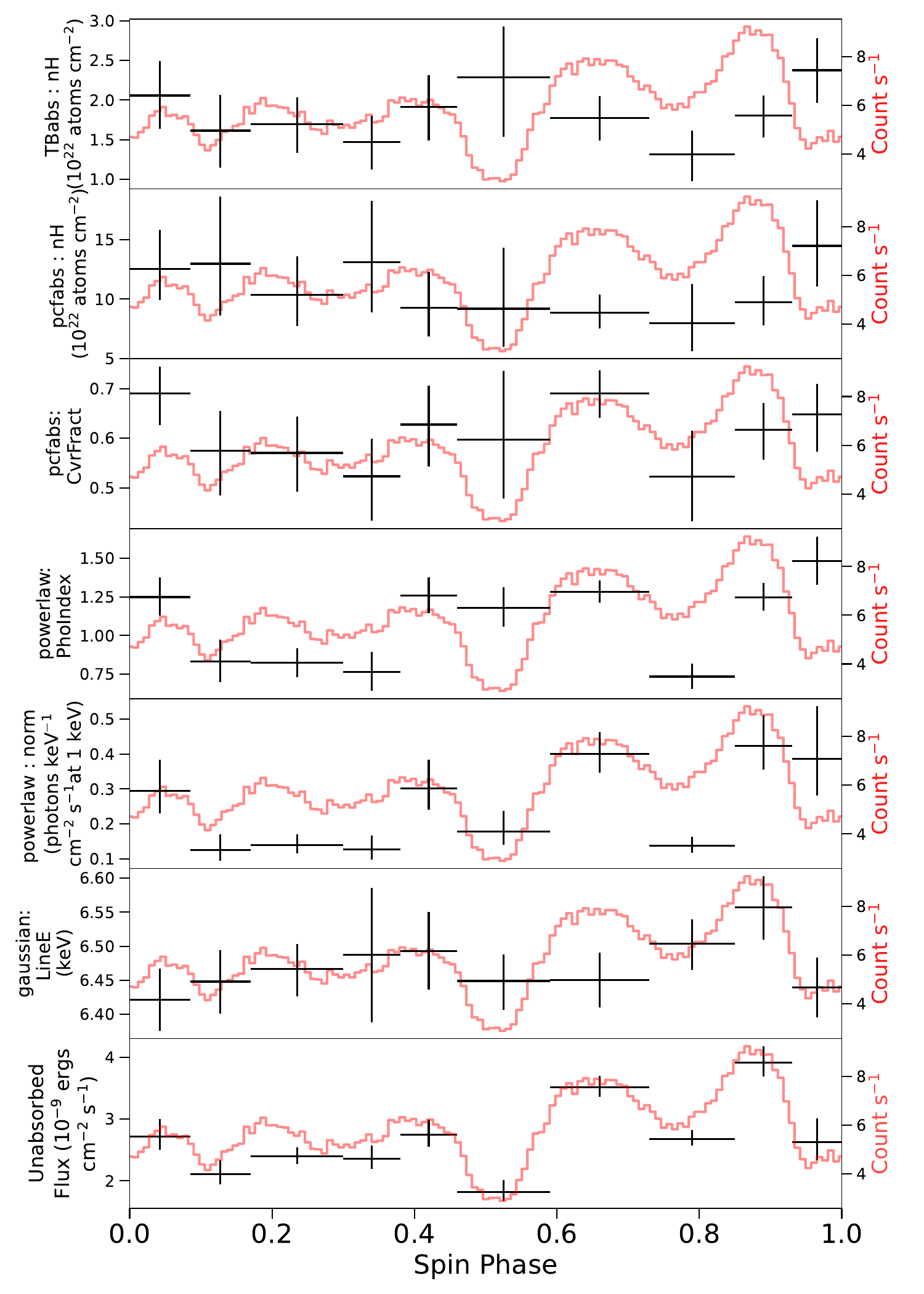}
    \caption{Variations of free spectral parameters plotted from the pulse-phase-resolved spectral analysis of the \text{0.9$-$9 $\mathrm{keV}$} XSPECT data for Vela X-1. The 0.9$-$9 $\mathrm{keV}$ unabsorbed flux for the spectral model used is calculated and plotted at the bottom. The errors are quoted at a 90\% confidence level.}
    \label{fig:phase_resolved_spectral_parameters} 
\end{figure}

We plot the parameter evolution with the spin phase in ${\text{Figure}\,\ref{fig:phase_resolved_spectral_parameters}}$ after fitting in the 0.9$-$9 keV energy range. We see the absorption rises near the pulse phase, $\phi_\text{spin}\sim0.5$ and $0.95$ for as seen for \texttt{tbabs:} nH. We also see a drop in absorption at $\phi_\text{spin}\sim0.8$ between the two peaks after $\phi_\text{spin}\sim0.5$. The partial absorption (\texttt{pcfabs:} nH) seems to quickly rise before $\phi_\text{spin}=1$.
Interestingly, we see that the Gaussian line center for the iron fluorescence emission has a quick rise and drop for the second peak at $\phi_\text{spin}\sim0.9$. This may be caused because of more ionized iron that has come into our line-of-sight and it was seen for the highest peak of the pulse profile. It should be noted such an ionized region is usually checked with multiple emission lines like the Fe XXV line at $\text{6.7 keV}$ and Fe XXVI line at 7.0 keV, along with the neutral 6.4 keV iron line. With current statistics, these lines if they are present cannot be resolved separately. Thus, requiring more data to analyze with improved statistics.

\subsection{Cyg X-1}

\begin{figure}
\includegraphics[width=0.48\textwidth]{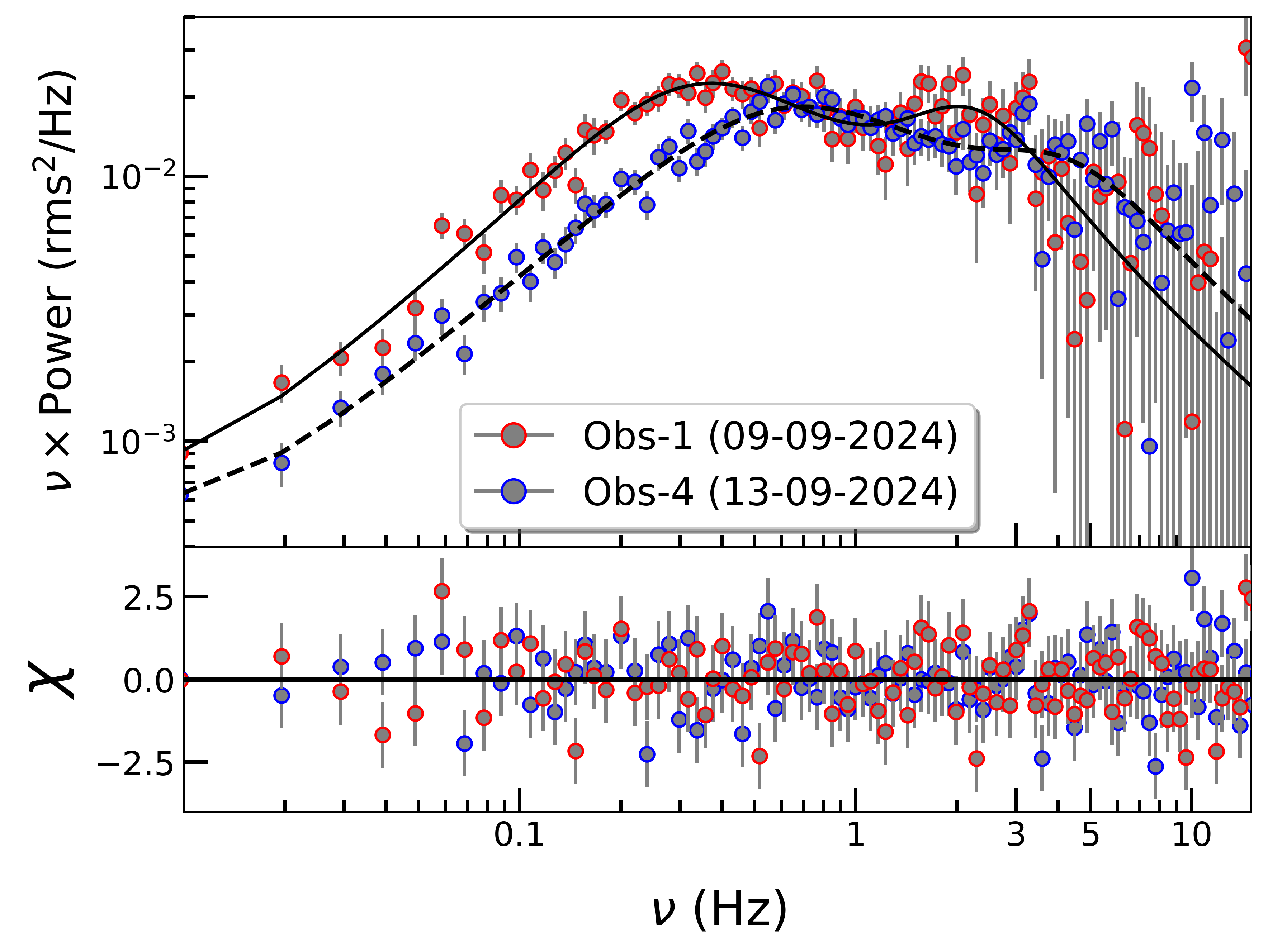}
\includegraphics[width=0.46\textwidth]{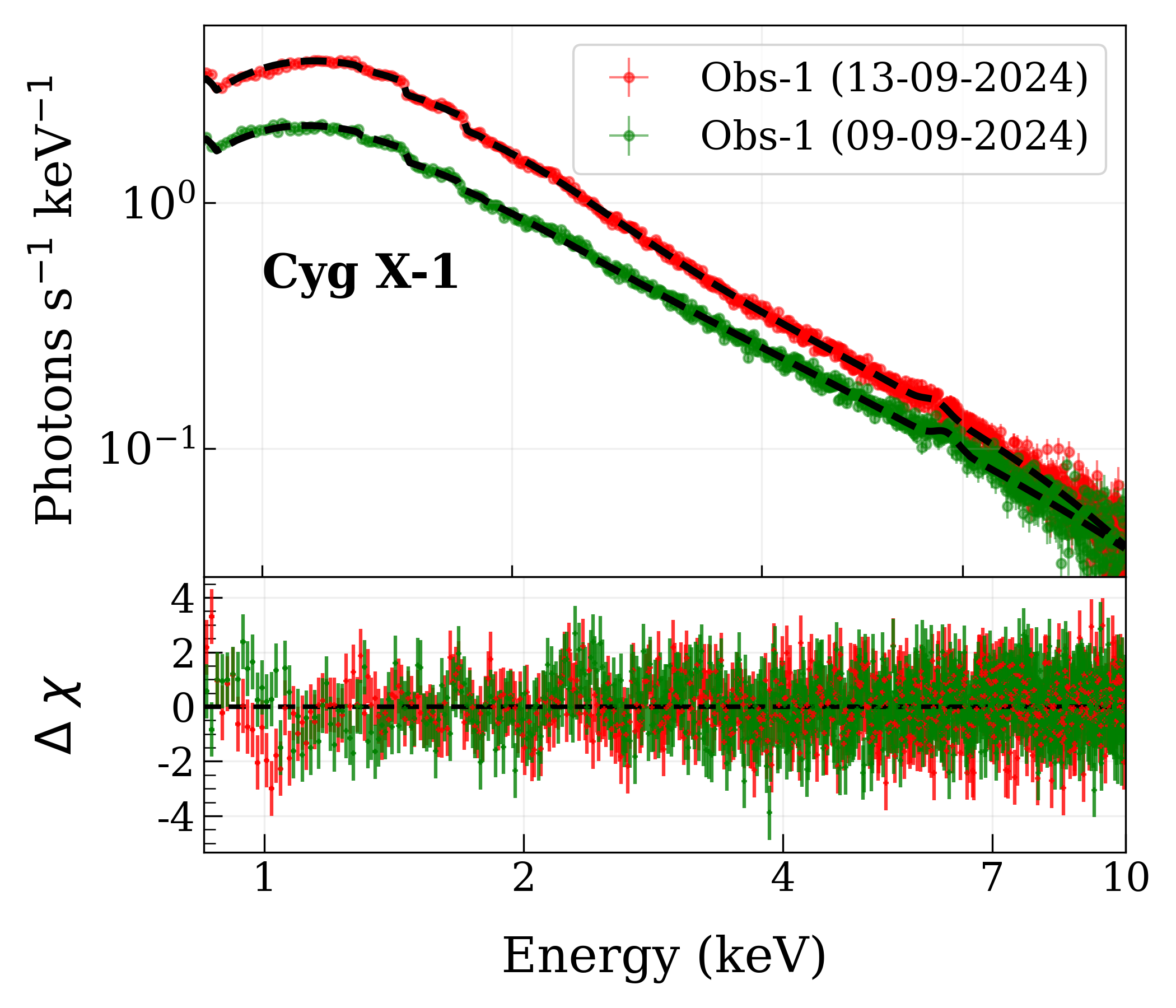}
 \caption{(Top) Wide-band PDS of Cyg X-1 on 9th September and 13th September 2024. (Bottom) The energy spectra corresponding to 9th September (green) and 13th September (red) 2024. Residuals are shown in bottom panels of each figure.}
 \label{CygX-1_spectra_figure}
\end{figure}

The four day long lightcurve shows a variation where, the count rate increases from $\sim$ 20 cts/s to $\sim$ 80 cts/s with time, whereas, hardness decreases from $\sim$ 0.35 to 0.2 as the time increases. This shows that the source transits towards a relatively softer state from MJD 60562 to MJD 60566.
The PDS are created for individual day-wise observations.
We fitted all the PDS using three \texttt{Lorentzian} functions (hereafter $L_0$, $L_1$ and $L_2$). 
The best-fit parameters for the PDS fits to four days of XSPECT observation are shown in Table \ref{tab1:PDS}.
The fitted PDS are displayed in the top panel of Figure \ref{CygX-1_spectra_figure}. The centroid frequency for one of the \texttt{Lorentzian} ($L_0$) is fixed at zero to represent the broad-band limited noise (BLN) \citep[see][for details]{ankur_2021_cygx1}. 
The PDS analysis implies a transition towards a relatively softer state as centroid frequency shifts towards a relatively higher frequency. 
No QPOs are detected in the PDS during the observations. 

\begin{table}
\renewcommand{\arraystretch}{1.4}
    \centering
    \caption{Details of the best fitted PDS parameters from \textit{XSPECT} observations of Cyg X-1 in $0.8-15$ keV energy range. The $L_1$ and $L_2$ corresponds to two broad \texttt{Lorentzian}s. All the errors are computed with 68\% confidence range.}
    \label{tab1:PDS}
    \resizebox{\columnwidth}{!}{
    \begin{tabular}{ccccc}
    \hline
    \hline
     Parameters & 9th Sept & 10th Sept & 12th Sept & 13th Sept \\
         \hline
     $\nu_1$ (Hz) & $0.16_{-0.03}^{+0.02}$ & $0.15_{-0.06}^{+0.06}$ & $0.21_{-0.05}^{+0.04}$ & $0.17_{-0.07}^{+0.06}$\\  
     $\sigma_1$ (Hz) & $0.63_{-0.08}^{+0.08}$ & $0.65_{-0.09}^{+0.10}$ & $0.81_{-0.12}^{+0.11}$ & $1.27_{-0.11}^{+0.13}$ \\  
      Norm$_{1}$ & $0.052_{-0.005}^{+0.004}$ & $0.045_{-0.006}^{+0.005}$ & $0.047_{-0.006}^{+0.004}$ & $0.048_{-0.002}^{+0.003}$ \\ 
        \hline
      $\nu_2$ (Hz) & $1.72_{-0.46}^{+0.30}$ & $1.02*$ & $2.20_{-1.61}^{+0.61}$ & $2.80*$ \\  
      $\sigma_2$ (Hz) & $2.65_{-0.68}^{+0.88}$ & $4.91_{-1.01}^{+1.51}$ & $4.35_{-1.54}^{+2.19}$ & $5.74_{-1.68}^{+2.57}$ \\  
     Norm$_{2}$ & $0.023_{-0.005}^{+0.006}$ & $0.030_{-0.005}^{+0.004}$ & $0.019_{-0.005}^{+0.006}$ & $0.014_{-0.003}^{+0.004}$\\ 
        \hline
     Total rms (\%) & $26.04\pm0.37$ & $25.93\pm0.34$ & $24.63\pm0.22$ & $23.96\pm0.26$ \\     
     $\chi^2$/dof & 120.25/105 & 134.04/106 & 120.59/105 & 103.24/106 \\
        \hline
        
    \end{tabular}
    }
    % \caption{Details of the best fitted PDS parameters from \textit{XSPECT} observations of Cyg X-1 in $0.8-15$ keV energy range. }
    % \label{tab1:PDS}
    \begin{list}{}{}
        \item[*] Frozen parameter.
	\end{list}
\end{table}

The model fitted spectra are shown in the bottom panel of Figure \ref{CygX-1_spectra_figure}. During the spectral fits, the hydrogen column density ($\text{n}_\text{H}$) was frozen at $0.60\times10^{22}$ atoms/cm$^{2}$ which agrees with the previous results (see \cite{ankur_2021_cygx1}). 
The inner disktemperature was observed to increase from 0.35 to 0.43 keV as the day progresses. The increase in inner disktemperature caused an increase in counts in lower energy band. Moreover, the spectral index ($\Gamma$) shows a decrease from 1.82 to 1.76. In addition, the electron temperature ($\text{kT}_e$) shows an overall decrease from 4.59 keV to 2.72 keV. Further, we estimated the bolometric luminosity of the source in the energy range $0.5-50$ keV in the unit of Eddington luminosity considering the mass and distance of the source to be 21.2 M$_{\odot}$ and 2.22 kpc respectively. The bolometric luminosity of the source was observed to increase gradually from 53\% to 80\% of $\text{L}_\text{Edd}$. 

\subsection{Cyg X-3}

The source lightcurve and HR trend from MAXI (see Figure \ref{CygX-3_lc_hr}) clearly show that Cyg X-3 is in the hard state before transitioning into an intermediate state. From XSPECT observations, it is found that the average HR in the $4-6$ keV band compared to $0.8-4.0$ keV is $0.93$ whereas the average HR in $6-10$ keV band compared to the $4-6$ keV band was found to be $0.69$. 
We do not detect any QPOs in our observations. 
The PDS is mainly dominated by weak red noise. We have fitted a power law function to the PDS. The power law index is constrained as $\sim 1.6-1.7$.

\begin{figure}
\centering
\includegraphics[width=0.48\textwidth]{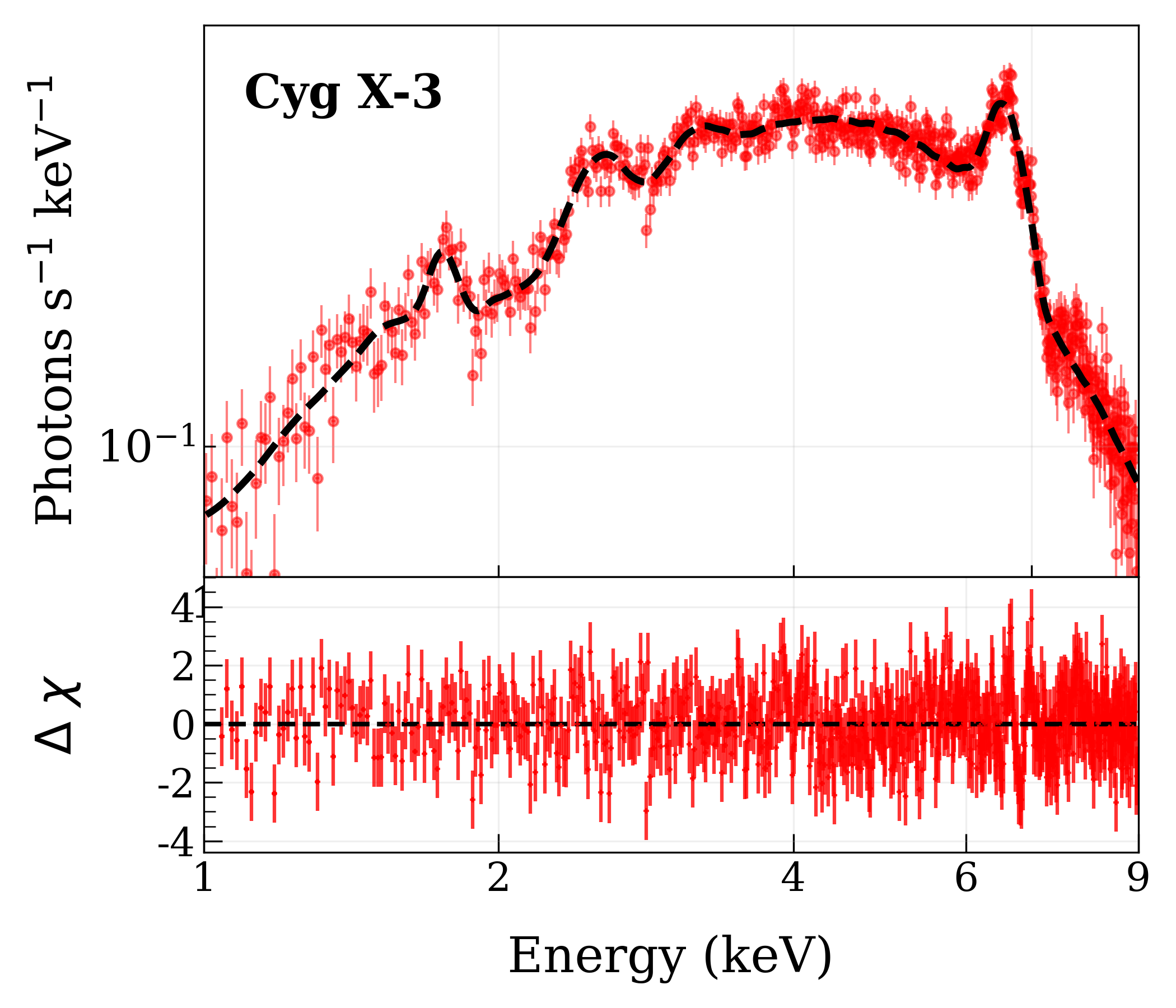}
 \caption{Time-averaged spectrum of Cyg X-3 in the energy band $1-9$ keV with the best-fit model shown with the black solid line: \texttt{tbabs*pcfabs(diskbb+Si+S XVI1+S XVI1+Fe)}. Four Gaussian lines represent the emission lines: Si: 1.94 keV, S XVI1: 3.15 keV, SXVI2: 2.51 keV, and Fe: 6.61 keV, respectively.}
 \label{CygX3_spectra_figure}
\end{figure}

The best-fit spectrum for the combined three days of observations is shown in Figure~\ref{CygX3_spectra_figure} and the best-fit parameters are given in Table~\ref{table_best_fit}. A combination of phenomenological models (\texttt{tbabs*(reflect*edge*thcomp*diskbb + 4*Gaussian)}) was used to analyze the spectrum obtained from simultaneous observations with \textit{AstroSat} (SXT and LAXPC) and Insight-HMXT while the source was in soft-state (\cite{2025ApJ...986...97C}). We used a simpler model combination: \texttt{tbabs*pcfabs*(diskbb + 4*Gaussian)}. Comparing common parameters values, the inner disktemperature obtained from our analysis is significantly larger ($\sim2.6$ keV) than what they have reported ($\sim0.89$ keV). Hence, it is understood that the source was in an intermediate-soft state during our observation. Our results also show there is a significant partial covering fraction ($\sim0.78$), implying the presence of clumpy stellar wind in the binary environment.

%\pagebreak
\begin{table*}
%\begin{sidewaystable}
\thispagestyle{empty}
\renewcommand{\arraystretch}{1.2}

    \caption{Best-fit spectral parameters for three categories of sources, NS-LMXBs, Pulsars and BH-XRBs. The analysis methodology and model description are described in Section \ref{ana_modeling} and \ref{red_and_disc}}
    \label{table_best_fit}
    \centering
    %\resizebox{\textheight}{!}{
    \begin{adjustbox}{max width=\textwidth}
        \begin{tabular}{cccccccccccc}   
    \hline
    \hline
    Segments & \makecell{$n_H$ \\ $\times10^{22}$ cm$^{-2}$} & $N_{H}$(pcfabs) & $f_{cov}$ & \makecell{$kT_{in}$ \\ (keV)} & $N_{dbb}$ & \makecell{$kT_e$ \\ (keV)} & $\Gamma$ & \makecell{$E_{Gauss-1}$ \\ (keV)} & \makecell{$E_{Gauss-2}$ \\ (keV)} & $\chi^2/dof$ & $L/L_{Edd}$ \\
    \hline
    \multicolumn{12}{c}{Aql X-1} \\
    \hline
     Persistent  & $ 3.57 \pm 0.05$ & --  & -- & $0.96 \pm 0.02$ & $628_{-30}^{+32}$ &  $1.67_{-0.03}^{+0.07}$ & $1.07_{-0.02}^{+0.26}$ & $6.66 \pm 0.03$ &  & 1.03  &   0.27 \\
    \hline
    \multicolumn{12}{c}{Sco X-1} \\
    \hline
     HB    & 0.14$\pm$ 0.004 & --  & 1(f) & 0.76$\pm$ 0.06 & 27280 $\pm$ 8292 & 2.82$\pm$ 0.07  & 1.70$\pm$0.02 & 6.71$\pm$0.02 & -- & 572/325 &  2.14  \\
     NB    & 0.12$\pm$ 0.003 &  -- & 1(f) & 0.85$\pm$ 0.02 & 19757 $\pm$ 2199 & 2.60$\pm$ 0.04  & 1.90$\pm$0.01 & 6.69$\pm$0.01  & -- &  378/321 & 1.89  \\
     FB    & 0.07$\pm$ 0.01  & --  & 1(f) & 0.62$\pm$ 0.03 & 37384 $\pm$ 8193 &  2.16$\pm$ 0.02 & 1.61$\pm$0.01 & 6.68$\pm$0.01 & -- &  368.9/325 & 2.42   \\
     \hline
     \multicolumn{12}{c}{Cir X-1} \\
     \hline
     EP  & 1.22(f)  & 4.41 $\pm$ 0.48 & 0.75 $\pm$ 0.05   & 0.56 $\pm$ 0.06 & -- & 3.93 $\pm$ 0.40 & 1.01 $\pm$ 0.16 & -- & -- &  1324/1139 &  0.041  \\
     ER  & 1.22(f) &  4.47 $\pm$ 0.23 & 0.95 $\pm$ 0.01  & 0.34 $\pm$ 0.08 & -- & 8.36 $\pm$ 0.97 & 1.33 $\pm$ 0.11  & 6.63 $\pm$ 0.20  & -- & 1302/1136  &  0.077  \\
     Tr   & 1.22(f) &  1.13 $\pm$ 0.47  & 0.56 $\pm$ 0.09 & 0.09 $\pm$ 0.01 & -- & 1.62 $\pm$ 0.08 & 1.45 $\pm$ 0.08 & -- & -- & 1169/1139 & 0.210  \\
     NeP  & 1.22(f) & 1.19 $\pm$ 0.20  & 0.95 $\pm$ 0.08  & 0.11 $\pm$ 0.01 & -- & 1.73 $\pm$ 0.04 & 1.60 $\pm$ 0.03 & -- & -- &  1840/1139 &  0.287  \\ 
     \hline
     \multicolumn{12}{c}{GX 301-2} \\
     \hline
     \makecell[c]{Day\\Averaged} & $15.90^{+3.24}_{-4.60}$ & $37.69_{-1.55}^{+1.65}$ & $0.89^{+0.02}_{-0.01}$ & -- & -- & -- & $0.95_{-0.05}^{+0.01}$ & $6.38_{-0.01}^{+0.03}$ & $7.01_{-0.04}^{+0.05}$ &  1854/1452 & $0.96$ \\
     \hline
     \multicolumn{12}{c}{Vela X-1} \\
     \hline
     Day-1  & 1.7$^{+0.3}_{-0.4}$ & 6.0$^{+1.4}_{-1.2}$  & 0.69$^{+0.07}_{-0.06}$ & -- & -- & 3.6$^{+0.6}_{-0.4}$ & -- & 6.47 $\pm\,$0.03 & -- & 1327/1443  & $0.83$ \\
     Day-2  & 1.5 $\pm\,$0.1  & 10.9$^{+1.7}_{-1.8}$ & 0.55$^{+0.05}_{-0.08}$ & -- & -- & 4.0$^{+2.1}_{-0.7}$ & -- & 6.49 $\pm\,$0.03 &  &  1356/1443 & $0.76$ \\
     Day-3  & 3.7$^{+0.8}_{-0.9}$  & 11.4$^{+3.7}_{-3.1}$ & 0.69$^{+0.08}_{-0.09}$ & -- & -- & 3.6$^{+1.6}_{-0.6}$ & -- & 6.46$^{+0.02}_{-0.03}$ & -- &  1341/1443 & $0.78$ \\
     \hline
    \multicolumn{12}{c}{Cyg X-1} \\
    \hline
     Day-1    & $0.6$(f) &  -- & -- & $0.35_{-0.01}^{+0.02}$ &  & $4.59_{-0.62}^{+1.26}$  & $1.82_{-0.02}^{+0.02}$ & -- & -- &  $835/804$ &  $0.53$ \\
     Day-2   & $0.6$(f) & --  & -- & $0.33_{-0.01}^{+0.01}$ &  & $4.71_{-0.67}^{+1.49}$  & $1.86_{-0.02}^{+0.02}$ & -- & -- &  $798/804$ &  $0.54$ \\
     Day-3    & $0.6$(f) & --  & -- & $0.39_{-0.01}^{+0.01}$ &  & $3.66_{-0.35}^{+0.52}$  & $1.85_{-0.03}^{+0.03}$ & -- & -- &  $771/804$ & $0.69$  \\
     Day-4    & $0.6$(f) & --  & -- & $0.43_{-0.01}^{+0.01}$ &  & $2.72_{-0.15}^{+0.19}$  & $1.76_{-0.04}^{+0.03}$ & -- & -- &  $835/804$ & $0.80$  \\    
    \hline
     \multicolumn{12}{c}{Cyg X-3} \\
    \hline
     Day-1  & 2.27$\pm0.43$ & 18.24$^{+3.67}_{-3.04}$ & 0.78$\pm0.05$ & 2.46$^{+0.16}_{-0.15}$ & 4.28$^{+1.26}_{-1.08}$ & -- & -- & -- & -- & 719/626  &  0.05 \\
     Day-2  & 2.46$\pm0.41$ & 25.99$^{+4.46}_{-3.84}$ & 0.81$\pm0.04$ & 2.71$^{+0.19}_{-0.17}$ & 7.33$^{+2.19}_{-1.91}$ & -- & -- & -- & -- & 652/626 & 0.06\\
     Day-3  & 2.33$^{+0.50}_{-0.49}$ & 20.73$^{+5.21}_{-3.98}$ & 0.78$^{+0.05}_{-0.06}$ & 2.69$^{+0.22}_{-0.19}$ & 4.22$^{+1.77}_{-1.24}$ & -- & -- & -- & -- & 709/626 & 0.05\\
    \hline
    \hline
    \end{tabular}
    %}
    \end{adjustbox}
%\end{sidewaystable}
\end{table*}

\section{Summary and Conclusion}

The XSPECT observations of the NS–LMXBs, namely, Aql X-1, Sco X-1, and Cir X-1 provide valuable insights into the diverse accretion and emission processes governing these systems. Aql X-1 exhibited persistent emission accompanied by a classical thermonuclear X-ray burst (type-I), confirming active accretion onto the neutron star surface and allowing constraints on the stellar mass and radius through spectral modeling (\cite{lewin1993},\cite{strohmayer2006}). Sco X-1, observed across its complete Z-track in the soft band for the first time with XSPECT, revealed systematic spectral variations along different branches, driven by accretion-rate fluctuations and changes in the disk–corona configuration. The consistent Fe K$\alpha$ detection across all states further supports the presence of a stable reprocessing region close to the neutron star. Cir X-1 demonstrated strong phase-dependent variability, transitioning from hard to soft states as the system moved from eclipse to non-eclipse phases, consistent with changing absorption and accretion geometry across the orbit (see \cite{tominaga2023}). Together, these results highlight the diagnostic power of XSPECT in probing the complex spectral evolution, thermonuclear activity, and accretion dynamics of NS–LMXBs, offering crucial constraints on neutron star structure and the physical mechanisms driving variability in these systems

The XSPECT observations of the accretion-powered pulsars GX~301$-$2 and Vela~X–1 provide important insights into their spin behavior, spectral evolution, and accretion geometries. In GX~301$-$2, a spin period of $671.4 \pm 0.1$~s was measured, consistent with measurements after the spin-up period in early 2019 (\cite{nabizadeh2019},\cite{2021MNRAS.506.2712D}). The pulse profile exhibited a double-peaked morphology and strong energy dependence, indicating complex emission geometry influenced by the accretion column and magnetic field configuration (\cite{2023hxga.book..138M},\cite{suleimanov2023}). The time-averaged spectrum, modeled with an absorbed power-law and iron fluorescence lines at $\sim$6.4 and 7~keV, revealed significant reprocessing in the stellar wind environment. A strong positive correlation between spectral hardness and intensity further suggests that spectral hardening accompanies enhanced accretion activity and localized absorption near the neutron star. In Vela~X–1, the photoelectric absorption was found to increase progressively with orbital phase ($\phi_\text{orb} \sim 0.15$–$0.48$), consistent with the neutron star traversing a dense accretion wake (\cite{malacaria2016},\cite{kretschmar2021}). 
The Comptonizing electron temperature remained nearly constant throughout, indicating a stable corona despite orbital evolution. Phase-resolved variations in spectral parameters around $\phi_\text{spin} \sim 0.9$ suggest possible changes in the ionization state of the accreting material, potentially linked to viewing geometry near the magnetic poles. 
Together, these results emphasize the complex interplay between wind accretion, magnetic field geometry, and orbital modulation in shaping the X-ray emission properties of classical high-mass X-ray pulsars.

BH-XRBs observations of XSPECT help probing the accretion behaviors and spectral evolution characteristic of their distinct environments. Cyg~X–1 exhibited a clear transition toward a softer spectral state between MJD~60562 and MJD~60566, as evidenced by an increase in count rate from 20 to 80~counts~s$^{-1}$ accompanied by a decrease in hardness ratio from 0.35 to 0.20. The broadband power density spectra displayed two broad Lorentzian components, with centroid frequencies shifting from 1.72~Hz to 2.80~Hz and a slight reduction in total fractional rms from 26.0\% to 24.0\%, indicative of enhanced disk contribution and reduced coronal variability. The observed increase in the disk temperature ($kT_{\mathrm{in}}$) and bolometric luminositNePy, along with a decrease in electron temperature ($kT_{e}$), further supports a state transition from the canonical hard to soft-intermediate state. In contrast, Cyg~X–3 remained in the intermediate state throughout the observation period, characterized by weak red noise in its power spectra and the absence of significant QPOs. The detection of multiple emission lines, including Fe at 6.61~keV, Si at 2.74~keV, and S~XVI lines near 2.5–3.2~keV, highlights reprocessing in a dense and clumpy stellar wind environment. These results underscore the ability of XSPECT to capture the spectral–timing diversity among BH–XRBs, tracing both intrinsic disk–corona evolution in Cyg~X–1 and the wind-dominated accretion dynamics in Cyg~X–3.

The first XSPECT/\textit{XPoSat} Workshop provided participants with invaluable hands-on exposure to XSPECT on-board science data, offering practical training in standard X-ray data analysis tools and techniques. Through guided sessions on calibration, data extraction, and reduction using standard pipelines, participants gained experience in generating Level-2 data products such as light curves and spectra. The hands-on exercises, complemented by detailed Python notebooks and short quizzes, deepened the understanding of XSPECT data processing and analysis. The workshop uniquely combined observational and instrument lectures with practical sessions, bridging the gap between instrumentation and science. Expert talks, delivered by scientists directly involved in the payload development, offered insightful perspectives on the scientific capabilities and functional aspects of the instrument, thereby enhancing participants’ comprehension of X-ray observational techniques and physics of compact objects.

A major highlight of the workshop was the collaborative group project, in which research scholars from across the country worked together under expert mentorship to analyze the initial XSPECT data from various category of sources (NS-LMXBs, Pulsars, BH-XRBs) and explore the underlying science of compact objects. This interactive and interdisciplinary engagement fostered teamwork, knowledge exchange, and the formation of potential future collaborations within the Indian astronomy community. The workshop’s emphasis on achieving tangible scientific outcomes, performing spectral and timing analysis rather than focusing solely on data extraction, made it both rigorous and productive. Such initiatives play a pivotal role in nurturing early research interest and strengthening expertise in X-ray astronomy, paving the way for future scientific explorations using XSPECT onboard \textit{\textit{XPoSat}} and forthcoming Indian space missions.

\section*{Data availability}
This publication uses the data from the \textit{XPoSat} mission of the Indian Space Research Organization (ISRO), archived at the Indian Space Science Data Centre 
(\href{https://www.issdc.gov.in/}{ISSDC}: \href{https://pradan1.issdc.gov.in/x01/}{https://pradan1.issdc.gov.in/x01/}).
%%Appendix

\section*{Acknowledgments}
Authors (participants) acknowledge support and guidance provided by XSPECT team with software and user manual to analyze the data during the workshop. The authors also thank \textit{XPoSat} project team, and mission team for their involvement and support in enabling XSPECT payload onboard \textit{XPoSat} mission. Special thanks to P. Sreekumar (Chairperson, SWG, \textit{XPoSat}) and S. Seetha (Chairperson, XTAC, \textit{XPoSat}) and the respective review committees for ensuring the high quality data products and availability of data to user community. The authors thank GD, SAG; DD, PDMSA and Director, URSC for encouragement and support to carry out this research. The \textit{XPoSat} project is managed and facilitated by the Indian Space Research Organization.

\bibliography{bibliography}
\bibliographystyle{apalike}

\end{document}